\documentclass[12pt]{article}
\usepackage{color,natbib,verbatim,graphicx}
\usepackage[affil-it]{authblk}
\usepackage{amsmath}
\usepackage{setspace}
\usepackage{caption}
\usepackage{subfig}
\usepackage{fancyhdr}
\usepackage{enumitem}
\usepackage{soul}
\usepackage{setspace}
\usepackage{caption}
\usepackage{subfig}
\usepackage{float}
\usepackage[noend]{algpseudocode}
\usepackage{algorithm}
\usepackage{setspace}
\usepackage{caption}
\usepackage{subfig}
\usepackage{amsmath, amssymb, amsfonts}

\algdef{SE}[SUBALG]{Indent}{EndIndent}{}{\algorithmicend\ }%
\algtext*{Indent}
\algtext*{EndIndent}
\def\R{\mathbb{R}}

\def\beqn{\begin{eqnarray*}}
\def\eeqn{\end{eqnarray*}}
\def\beq{\begin{eqnarray}}
\def\eeq{\end{eqnarray}}

\def\ld{\ldots}
\def\bi{\begin{itemize}}
\def\ei{\end{itemize}}
\def\ra{\rightarrow}

\def\X{\mbox{\boldmath $X$}}
\def\Z{\mbox{\boldmath $Z$}}
\def\Y{\mbox{\boldmath $Y$}}

\def\x{\mbox{\boldmath $x$}}

\def\cv{cross-validation}

\definecolor{mygreen}{rgb}{0.1,0.6,0.0}

\definecolor{myblue}{cmyk}{0.95,0.5,0,0.3}
\definecolor{mygreen}{cmyk}{1,0,0,0.74}
\definecolor{mygreen2}{rgb}{0.1,0.6,0.0}
\definecolor{myred}{rgb}{0.9,0,0.2}
\definecolor{purp}{rgb}{0.35,0,0.65}
\def\tcb#1{\textcolor{blue}{#1}}
\def\tcr#1{\textcolor{myred}{#1}}
\def\tcp#1{\textcolor{purp}{#1}}

{\markright\today}

\begin{document}

\title{\bf Use of Cross-validation Bayes Factors to Test Equality of
  Two Densities}
\author{Naveed Merchant\footnote{Department of Statistics, Texas A\&M University, {\tt nmerchant@stat.tamu.edu}},\, Jeffrey D.~Hart\footnote{Department of Statistics, Texas A\&M University, {\tt hart@stat.tamu.edu}}\,\, and Taeryon Choi\footnote{Department of Statistics, Korea University, {\tt trchoi@korea.ac.kr}}}

\date{}
\maketitle

\begin{abstract}
We propose  a non-parametric, two-sample Bayesian test for checking whether or not two data sets share a common distribution. The test makes use of data splitting ideas and does not require priors for high-dimensional parameter vectors as do other nonparametric Bayesian procedures. We provide evidence that the new procedure provides more stable Bayes factors than do methods based on P\'olya trees. Somewhat surprisingly, the behavior of the proposed Bayes factors when the two distributions are the same is usually superior to that of P\'olya tree Bayes factors. We showcase the effectiveness of the test by proving its consistency, conducting a simulation study and applying the test to Higgs boson data. 

\end{abstract}

\section{Introduction}\label{intro}
In frequentist hypothesis testing, there is no universal statistic whose values are interpretable across different problems.  In contrast, Bayes factors {\it do} have a universal interpretation. When the prior probabilities of two hypotheses are the same, the Bayes factor is the ratio of posterior probabilities of the two hypotheses. This is a compelling motivation for developing objective Bayesian procedures that depend only minimally on prior distributions. 
\cite{HC} proposed the use of {\it cross-validation} Bayes factors (CVBFs) 
to compare the fit of parametric and nonparametric models.  The CVBF is an objective Bayesian procedure in which the 
nonparametric model is a kernel density estimate, the 
simplest version of which cannot typically be used in a Bayesian
analysis  since it only becomes a model once it is computed from
data. This problem is sidestepped by computing a kernel estimate
from a subset of the data, and then using the estimate as a model for the
remainder of the data.  
As detailed by \cite{HM}, the notion of a CVBF is also useful in a
purely parametric context, wherein data splitting allows one to
compare two parametric models via a legitimate Bayes factor that does not require a prior  distribution for either model. 
 
The purpose of the current paper is to explore CVBFs in the problem of
comparing densities corresponding to two different populations. Given
independent random samples $X_1,\ld,X_m$ and $Y_1,\ld,Y_n$ from
densities $f$ and $g$, respectively, we wish to test the
null hypothesis that $f$ and $g$ are identical against the
alternative that they are different, {\it without specifying a
parametric model for either density.} This is accomplished by means of
a Bayes factor that makes use of data splitting and kernel density
estimates. Unlike the setting of either \cite{HC} or \cite{HM}, both
hypotheses in the current setting are nonparametric, which
necessitates different techniques to show that CVBFs behave
desirably. In particular, it is of interest to prove that a CVBF is
Bayes consistent when either $f\equiv g$ or the two densities are
different. Although the current investigation is restricted to
comparison of two densities, we will lay the groundwork for justifying
the use of CVBFs in other settings where both hypotheses are
nonparametric. 

A classic Bayesian approach for checking the 
equality of two densities involves the construction of priors
on the elements of a wide class of
distributions. For testing goodness of fit and obtaining posterior predictive distributions, \cite{hanson2006inference} proposes methodology 
based on a P\'olya tree prior constructed from 
a centering distribution. 
Methods that use a similar strategy for checking
equality of two densities have been suggested 
by \cite{wong2010optional}, \cite{chen2014bayesian} and
\cite{holmes2015two}. \cite{dunson2008bayesian} propose the use of
  restricted dependent Dirichlet process priors when testing the
  equality of distributions against ordered alternatives. Both \cite{holmes2015two} and \cite{chen2014bayesian} use their Bayes factors in frequentist fashion, i.e., they choose rejection regions to produce desired type I error probabilities. In our opinion, such an approach is not truly Bayesian. If one uses a traditional level of significance such as 0.05, this practice yields a test with the unsettling property that in some cases the hypothesis of equal densities is rejected when the Bayes factor {\it favors} equal densities. We prefer an approach that chooses the hypothesis of unequal densities only when the odds in favor of unequal densities has increased in light of the data.

In Section
\ref{cons} it will be seen that any two-sample procedure 
based on Bayes factors ultimately depends on the difference between
entropy estimates.   
\cite{beirlant1997nonparametric}  have suggested using either a kernel
density estimate or histogram to estimate entropy and provide
conditions under which these types of estimators are 
consistent. In addition, \cite{beirlant1997nonparametric} use 
entropy estimates to check equality of distributions. 
 Entropy estimates are also seen in
 the information gain filter in machine learning methods for 
 picking important features; see \cite{sarkar2013empirical}. The current paper makes use of results in \cite{Hall87}, who proves consistency of entropy estimates that rely on data-driven smoothing parameters.

An important issue is that of Bayes factor consistency, which we address in
Section \ref{cons}.  Suppose the Bayes factor is defined so that values smaller than 1 favor the hypothesis of equal densities. Then consistency means that the Bayes factor converges in probability to $0$ when the densities are equal and diverges to $\infty$ when the densities are unequal. \cite{holmes2015two} 
contains a proof showing that their Bayes factor is consistent. We argue that our cross-validation Bayes factor is consistent as well. Moreover, when the densities are equal, we argue that a cross-validation Bayes factor converges to $0$ at a much faster rate than do Bayes factors based on traditional Bayesian methods. 

A main motivation for our proposed methodology is its conceptual
simplicity. The models used are kernel density estimates from training
data and each one depends on but a single parameter, a bandwidth. In
contrast, the approaches 
of \cite{hanson2006inference} and \cite{holmes2015two} depend on
choice of base distribution and $2^k$ parameters, where $k$ is
typically at least 10. One also needs to choose a prior for all these
parameters, although \cite{holmes2015two} propose one that
requires specification of just one parameter. In simulations 
in Section \ref{sims} we will compare our method with that of
\cite{holmes2015two}, and show that the odds ratios produced by the
latter test can be highly sensitive to the choice of base distribution. 

The rest of the paper may be outlined as follows. In Section
\ref{method} we describe in detail our methodology for the two-sample
problem. Section \ref{kp} considers the choice of kernel and also the
prior used for the bandwidth parameter, and Section \ref{lapapp}
investigates the use of a Laplace approximation for marginal
likelihoods. In Section \ref{cons} we provide theoretical evidence
that our Bayes factor is consistent, and in Section \ref{tv} we
discuss methods for choosing the training set sizes. Finally, Sections
7 and 8 are devoted to a simulation study and real-data analysis,
respectively. 

\section{Methodology}\label{method}

Suppose that we observe independent random samples $X_1,\ld,X_m$ and
$Y_1,\ld,Y_n$ from cumulative distribution functions $F$ and $G$,
respectively. We assume that $F$ and
$G$ have respective densities $f$ and $g$, and the goal is to test the
following hypotheses by means of a Bayesian approach:
$$
H_0:\, f\equiv g\quad{\rm vs.}\quad H_a:\, f\not\equiv g.
$$
We wish to use kernel density estimates to do the testing, and in
order to do so  we will use the CVBF idea. In contrast to the setting
of  \citet{HC}, both the 
null and alternative hypotheses are nonparametric, and hence training data will
be used to formulate the alternative {\it and} null models. The Bayes
factor will then be computed from validation data. 

We first introduce some notation. For an arbitrary collection of
  (scalar) observations $\Z=(Z_1,\ld,Z_n)$, define the kernel density
  estimate (KDE) $\hat f(\,\cdot\,|h,\Z)$ by
$$
\hat f(x|h,\Z)=\frac{1}{nh}\sum_{i=1}^nK\left(\frac{x-Z_i}{h}\right),
$$
where the kernel $K$ is a probability density and $h>0$ is the
bandwidth. For the moment, all we ask of $K$ is that it be symmetric
about 0, unimodal and have finite variance.

Now, partition $X_1,\ld,X_m$ into $\X_T=(X_1,\ld,X_r)$ and
$\X_V=(X_{r+1},\ld,X_m)$, 
and likewise $Y_1,\ld,Y_n$ into $\Y_T=(Y_1,\ld,Y_s)$ 
and $\Y_V=(Y_{s+1},\ld,Y_n)$. Under $H_0$ there is a common density,
call it $f$.  The 
model for $f$ will be $M_0=\{\hat f(\,\cdot\,|h,\X_T,\Y_T):h>0\}$. In
other words, we pool the two training sets together and use these data
to estimate the common density $f$.
Under the alternative we have separate models for $f$ and
$g$, which are $M_X=\{\hat
f(\,\cdot\,|\alpha,\X_T):\alpha>0\}$ and $M_Y=\{\hat
f(\,\cdot\,|\beta,\Y_T):\beta>0\}$. 

Let $\pi$, $\pi_X$ and $\pi_Y$ be priors for $h$, $\alpha$ and
$\beta$, respectively.  The likelihood under $H_0$ is 
$$
L_0(h)=\prod_{i=r+1}^m\hat f(X_i|h,\X_T,\Y_T)\prod_{j=s+1}^n\hat
f(Y_j|h,\X_T,\Y_T).
$$
The likelihood under $H_a$ is
$$
L_a(\alpha,\beta)=\prod_{i=r+1}^m\hat f(X_i|\alpha,\X_T)\prod_{j=s+1}^n\hat
f(Y_j|\beta,\Y_T)=L_X(\alpha)L_Y(\beta),
$$
and the \cv\ Bayes factor (CVBF) is
\beq\label{CVBF}
CVBF&=&\frac{\int_0^\infty\int_0^\infty\pi_X(\alpha)\pi_Y(\beta)L_a(\alpha,\beta)\,d\alpha  
d\beta}{\int_0^\infty\pi(h)L_0(h)\,dh}\notag\\ 
&=&\frac{\int_0^\infty\pi_X(\alpha)L_X(\alpha)\,d\alpha\cdot
  \int_0^\infty\pi_Y(\beta)L_Y(\beta)\,d\beta}{\int_0^\infty\pi(h)L_0(h)\,dh}. 
\eeq

Interestingly, each of $M_0$,
$M_X$ and $M_Y$ is a parametric model, inasmuch as each depends on 
just a single parameter, a bandwidth. It should be acknowledged that
we know with certainty that, for example, $M_X$ {\it does not} contain
the true 
density $f$. However, there is a key difference between $M_X$ and a
traditional one-parameter model. Since KDEs are consistent
estimators, we have reason to believe that some members of
$M_X$ will be quite close to $f$, especially if the training set
size $r$ is large. In contrast, members of a traditional one-parameter
model, such as all $N(\mu,1)$ densities, would be close to the truth
only under very special circumstances. So, in spite of being formally
``wrong,'' $M_X$ can be expected to be a good model, which echoes the
sentiment of George Box in his famous quote about statistical
models. 

Even if one objects to our models not formally containing the truth,
the same criticism can arguably be leveled against the P\'olya tree
approach of \cite{holmes2015two}. Each element of the parameter space
in that approach is of histogram type, and since one usually envisions
a certain degree of smoothness in the underlying density, the true
density does not necessarily lie in the parameter space employed by
P\'olya trees.

We close this section with some remarks about our methodology. 
\bi
\item The quantity (\ref{CVBF}) is referred to as a \cv\ Bayes factor \citep{HC}
  since each data set is split into two parts. For example, the data
  $X_1,\ld,X_m$ are split into a training set, $\X_T$, and a
  validation set, $\X_V$. 
\item In spite of the fact that the models being compared in $CVBF$
  are formulated from data, it is important to appreciate that $CVBF$
  {\it is} a legitimate Bayes factor.  This is because the models are
  defined from data that are independent of the validation sets
  $\X_V$ and $\Y_V$. The Bayesian paradigm does not specify
  {\it where} posited models must come from, so long as they
  are not defined from the data used to evaluate those models.
\item By assuming that the bandwidths $\alpha$ and $\beta$ are a
  priori independent, the computation of $CVBF$ reduces to calculating
  three separate marginals, each of which has the form dealt with in
  \cite{HC}.
\item \cite{CIS-5329} show that the $L_1$ norm difference between a
  kernel density estimate and the true density tends to 0 
 for any kernel integrating to 1 as long as the sample size $n$ tends to
 $\infty$, the bandwidth $h$ tends to 0, and $nh\ra\infty$. Because of
 results like this, the conventional wisdom in kernel density
 estimation is that the 
  choice of kernel $K$ is not overly important.  This is not at all
  the case in the current context. In Section \ref{kp} we
  will point out the importance of using relatively {\it heavy-tailed}
  kernels, a specific version of which is proposed. 
\item Ideally a CVBF should not depend on the particular data split
  that is used. Therefore, we suggest that one use the
  geometric mean of $CVBF$ values calculated from a number of
  different randomly chosen splits.  
\ei

\section{Implementation issues}
Some practical issues must be addressed in order to make use of CVBFs. A kernel has to be chosen for each of the KDEs, and priors for the bandwidths of the KDEs are needed. Furthermore, the integrals defining the three marginals cannot (in general) be computed analytically, and hence approximations of the integrals are necessary. We first address the choice of kernel and priors.

\subsection{Choice of kernel and priors}\label{kp}

For densities $f_1$ and $f_2$, the Kullback-Leibler divergence between
$f_1$ and $f_2$ is defined to be
$$
KL(f_1,f_2)=\int_{-\infty}^\infty f_1(x)\log\left(\frac{f_1(x)}{f_2(x)}\right)\,dx.
$$
As will be discussed in Section \ref{cons}, consistency of our
proposed Bayes factor depends crucially on the behavior of $KL(f,\hat
f(\,\cdot\,|\alpha,\X_T))$ and $KL(g,\hat
f(\,\cdot\,|\beta,\Y_T))$. \cite{Hall87} shows that the right sort of
kernel needs to be used to ensure that these 
divergences are well-behaved. A number of practical and technical
difficulties arising from tail behavior of the underlying density are
eliminated if one uses a relatively heavy-tailed kernel. A kernel that
suffices in this regard is the following that was proposed by \cite{Hall87}: 
\beq\label{Hallkern}
K_0(z) = \frac{1}{\sqrt{8 \pi e}\,\Phi(1)}\exp\left[-\frac{1}{2}(\log(1+|z|))^2\right],
\eeq
where $\Phi$ is the standard normal distribution function. 

\cite{Hall87} provides an example of when the popular Gaussian kernel
can fail in our context. Suppose that $f$ is a Cauchy density, and one
estimates $f$ by a kernel estimator $\hat f_h$ with Gaussian kernel
and bandwidth $h$. Then (i) the
expected Kullback-Leibler loss of $\hat f_h$ is infinite and (ii)
the likelihood cross-validation choice of $h$ diverges to infinity. 
In contrast, if kernel $K_0$ is used in this case, then the likelihood
cross-validation bandwidth is asymptotic to the minimizer of expected
Kullback-Leibler loss.  Simulations in Section \ref{sims} show that 
  these results for the Cauchy distribution appear to be true for the 
  version of \cv\ used in the current paper.

If one is confident that the tails of the underlying density are
no heavier than those of a Gaussian density, then it would be
appropriate to use a Gaussian kernel in our procedure. Simulations we
have done suggest that the Gaussian kernel produces somewhat more
stable Bayes factors than does $K_0$ in the case of light-tailed
densities. However, when there is uncertainty about the tails of the
underlying density, $K_0$ is a much better 
choice for the kernel. For this reason we will use $K_0$ for all 
simulations and data analyses in this paper.

The prior we propose for each bandwidth is as follows, which is the same type as
used by \cite{HC}:
\beq\label{prior}
\pi(h|\gamma) =
\frac{2\gamma}{\sqrt{\pi}h^2}\exp\left(-\frac{\gamma^2}{h^2}\right)I_{(0,\infty)}(h). 
\eeq
An aspect of this prior that we find appealing is that it
tends to $0$ as $h$ tends to 0. This in concert with the fact that,
due to the data-driven nature of 
our kernel density estimation models, we are (essentially) a
priori certain that the very smallest bandwidths produce
untenable densities. 

The mode of (\ref{prior}) is $\gamma$, and we propose that for each 
marginal, $\gamma$ be chosen to equal the maximizer of the corresponding
likelihood. For example, for the 
marginal $\int_0^\infty\pi_X(\alpha)L_X(\alpha)\,d\alpha$, we take
$\pi_X\equiv \pi(\,\cdot\,|\hat\gamma)$, where $\hat\gamma$ is the
maximizer of $L_X$. The scale of $\pi(\,\cdot\,|\gamma)$ is proportional
to $\gamma$, which entails that the prior $\pi(\,\cdot\,|\hat\gamma)$ has
low information relative to the likelihood. This is because the
variance of the \cv\ bandwidth $\hat\gamma$ is $o(\hat\gamma)$, a fact
that is ensured by using the kernel $K_0$. Centering a low information
prior at the maximizer of the likelihood is akin 
to using a unit reference prior \citep{consonni2018prior} centered at the data,
which by now is a fairly common practice. 

\subsection{Laplace approximation}\label{lapapp}

Interestingly, there exists a closed-form expression for each marginal
{\it if} one uses a Gaussian kernel in conjunction with a prior of the
form (\ref{prior}). This results from the fact that, for example,
$\pi(\alpha|\gamma)L_X(\alpha)$ is a linear combination 
of functions each of which is proportional to a function of the form
$\alpha^k\exp(-A/\alpha^2)$, whose integral over $(0,\infty)$ may be
expressed in terms of the gamma function. The practical usefulness of
this closed-form solution is limited for two reasons. First, as was
noted in Section \ref{kp}, the Gaussian kernel is not a good
all-purpose kernel, and secondly it turns out that 
more computations are required for the closed form solution than for
standard methods of approximating integrals. The solution requires
$r^{m-r}$ sums to be computed, where $r$ is the size of the training
set and $m-r$ the 
size of the validation set.  For these reasons we will not 
pursue the closed form solution further.

In general, the integrations required to calculate a $CVBF$ cannot be done
analytically. \cite{HC} used numerical integration to approximate
marginal likelihoods, either by simple or adaptive quadrature.
Other methods that could be used are 
importance sampling, bridge sampling or a Laplace approximation. 
A Laplace approximation has the advantage of being less computationally
intensive. Let $\hat h$ be the maximizer of $L_0$ and define 
\[
\hat H=-\frac{\partial^2}{\partial h^2}\log L_0(h)\Big|_{h=\hat h}.
\]
Then the Laplace approximation of $\int \pi(h)L_0(h)\,dh$ is
\[
\int \pi(h)L_0(h)\,dh\approx \sqrt{\frac{2\pi}{\hat H}}\cdot\pi(\hat h)L_0(\hat h).
\]
The quantity $\hat H$ can be expressed as a functional of kernel
estimates based on the kernel $K$ and two other related kernels. An
expression for $\hat H$ may be found in the Appendix. 

To investigate how well the Laplace approximation works in our context
we generate samples from a standard normal distribution and compare
the Laplace approximation of the marginal likelihood with an
approximation using the R function {\tt integrate} (which uses adaptive quadrature).  Data were
generated from a standard normal distribution and sample sizes 200,
500 and 1000 were considered. The kernel used was $K_0$, and the prior
was (\ref{prior}) with $\gamma$ taken to be the
maximizer of the likelihood. The training set size was always 1/4 of
the sample size, and 500 replications for each $n$ were
considered. Table 1 summarizes the results. 

\begin{figure}[t]

\begin{center}

{\includegraphics[height=3in]{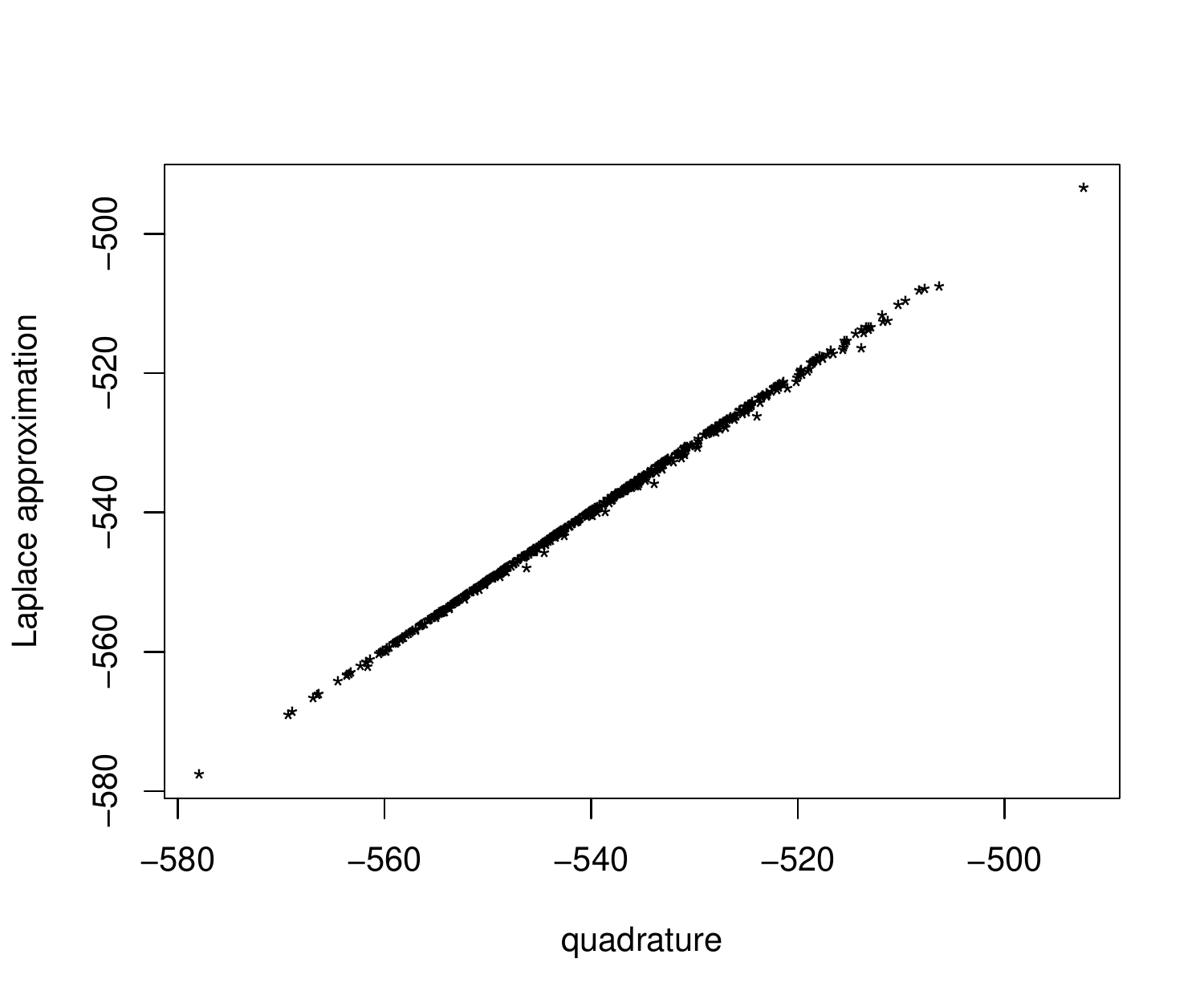}}

\caption{\it Laplace and quadrature approximations to log-marginal
 likelihoods. These results are for the case where the training set
 and validation sizes were $125$ and $375$, respectively.} 

\label{lapvsquad}

\end{center}

\end{figure}

\begin{table}[!htbp]

\Large


\begin{center}

\scalebox{0.70}{              

\begin{tabular}{|r|c|c|}


\hline

$n$ & Median & Interquartile range\\

\hline

200     & $6.99\cdot 10^{-4}$ &  $3.46\cdot 10^{-4}$ \\

\hline

500     & $2.66\cdot 10^{-4}$ & $1.80\cdot 10^{-4}$ \\

\hline

1000    & $1.33\cdot 10^{-4}$ & $2.77\cdot 10^{-4}$\\

\hline

\end{tabular}


} 

\end{center}

\caption{ {\it Relative error of Laplace approximation of marginal
    likelihood. Each median and interquartile range is based on 500
    replications. The measure of error is $|(log\hat M-\log M)/\log
    M|$, where $M$ and $\hat M$ are quadrature and Laplace
    approximations, respectively, of the marginal.}  } 

\label{tab:Laperror}

\end{table}
The Laplace approximations were excellent, with the median relative
error being no larger than 0.000699 and becoming smaller as the sample
size increased. A plot of the results for $n=500$ is given in Figure
\ref{lapvsquad}. We also found that the computations for our Laplace
approximation are 7 to 8 times faster at $n=1000$ than those for the
quadrature approximation when running on an 8 core Intel Skylake
  6132 CPU running at 2.6GHz with 32GB of 2666MHz DDR4 memory. For
these reasons we will use the Laplace approximation in all subsequent
simulations and examples. 

We note that the parameter of our prior is chosen so that the
  prior mode is equal to the maximizer of the likelihood. This has two
  computational benefits. First of all, our algorithm starts by
  determining the 
  maximizer of the log-likelihood, which is necessary to avoid
  underflow problems.  But once this maximizer has been determined it
  is not necessary to find the posterior mode since the two
  quantities are one and the same. Secondly, choosing the prior
  parameter in this way renders null the
  distinction between the two versions of 
the Laplace approximation, one using the maximizer of $L_0$ and the
other the maximizer of $\pi L_0$.

\section{Bayes consistency}\label{cons}

Here we address the consistency of a $CVBF$ in the two-sample
problem. We begin with a list of assumptions.
\bi

\item[A1.] The Laplace approximation of each of the three marginals is
  asymptotically correct in that the log of the marginal likelihood is
  equal to the log of the Laplace approximation plus a term that is
  negligible in probability relative to the approximation.

\item[A2.] The densities $f$ and $g$ are bounded away from 0 and
  $\infty$ on    
$(-\lambda, \lambda)$ for each $\lambda > 0$, with 
$f(x) \sim c_1 x^{-a_1}$ and $f(-x) \sim c_2 x^{-a_2}$ as
  $x\ra\infty$, where both $c_1$ and $c_2$ are positive and $a_1$ and
  $a_2$ larger than $1$. Density $g$ satisfies the same properties as
  $f$, albeit with possibly different constants.

\item[A3.] The second derivatives $f''$ and $g''$ exist and are bounded and  
  almost   everywhere continuous on $(-\infty, \infty)$. In
  addition,  for a constant $C_2 < \infty$, 
$$
|f''(x)|\leq C_2 x^{-a_1 - 2}\quad {\rm and} \quad |f''(-x)| \leq C_2
x^{-a_2 - 2}\quad{\rm for}\ x>1,
$$
where $a_1$ and $a_2$ are the same as in A2. The function $g''$
satisfies the same properties as $f''$ with possibly different
constants.

\item[A4.] The kernel used is $K_0$, as defined in (\ref{Hallkern}). 

\ei
Conditions A2 and A3 are those of \cite{Hall87} and are needed to
ensure that the maximizer of the likelihood cross-validation criterion
is optimal in a Kullback-Leibler sense. \cite{Hall87} provides
another set of conditions that could be used in place of A2 and
A3. These conditions deal with compactly supported densities, but for the
sake of brevity we do not repeat these. Furthermore, the kernel
  conditions of \cite{Hall87} are satisfied by kernels other than
  $K_0$, but notably the Gaussian kernel does not satisfy
  his conditions.

We wish to argue that the 
  statistic $CVBF$ defined by (\ref{CVBF}) is Bayes consistent under
  both null and alternative hypotheses. Initially, we will consider
  the null case, i.e., $f\equiv g$.  Each Laplace approximation
  depends on a term $\widehat H$ for which $\log\widehat H=O_p(\log
  n)$. Define 
$$
\widehat{BF}=\frac{\pi_X(\hat\alpha)L_X(\hat\alpha)\pi_Y(\hat\beta)L_Y(\hat\beta)}{\pi(\hat h)L_0(\hat h)}, 
$$
where $\hat\alpha$, $\hat\beta$ and $\hat h$ are the maximizers of
$L_X$, $L_Y$ and $L_0$, respectively. We will see that $\log n$ is of
smaller order than $\log\widehat{BF}$ under both hypotheses, and
therefore by A1 it is sufficient to consider $\widehat{BF}$ when
investigating consistency. 

Although probably not necessary, we assume at this point that $m-r>r$ and $n-s>s$.
An important aspect of $\widehat{BF}$ is the behavior
of the maximizers of the likelihoods. Consider, for example,
$\hat\alpha$. We claim that under 
general conditions $\hat\alpha$ is asymptotic in probability to the minimizer,
$\alpha_r$, of $E[KL(f,\hat f(\,\cdot\,|\alpha,\X_T))]$ as $r$ 
and $m-r$ tend to $\infty$. \cite{Hall87} proved precisely
this result assuming that the KDE uses kernel (\ref{Hallkern})  
and its bandwidth is chosen by {\it leave-one-out}
likelihood cross-validation. This version of cross-validation chooses the bandwidth of $\hat f(\,\cdot\,|\alpha,\X_T)$ to maximize, with respect to $\alpha$,
\begin{equation}\label{loo}
\ell(\alpha)=\prod_{i=1}^r\hat f(X_i|\alpha,\X_T^i),
\end{equation}
where $\X_T^i$ is all the training data except for $X_i$, $i=1,\ld,r$. 
Using essentially the same proof as in \cite{Hall87}, it can be shown that our version of likelihood cross-validation is at least as efficient as the leave-one-out version.  Intuitively, this is plausible since our version of the cross-validation curve is 
\begin{equation}\label{ourversion}
L(\alpha)=\prod_{i=r+1}^{m}\hat f(X_i|\alpha,\X_T).
\end{equation}
In comparison to (\ref{loo}), (\ref{ourversion}) has two advantages: $m-r>r$ and the validation data $X_{r+1},\ld,X_m$ are completely independent of $\hat f(\,\cdot\,|\alpha,\X_T)$.
It 
  is also worth noting that \cite{LDK} prove optimality of  
  the version of likelihood cross-validation that we use, albeit under
  more restrictive conditions than those of \cite{Hall87}.

Now, let $\beta_s$ and $h_{r+s}$ denote the minimizers of
  $E[KL(f,\hat f(\,\cdot\,|\beta,\Y_T))]$ and $E[KL(f,\hat
    f(\,\cdot\,|$ $h,\X_T,\Y_T))]$, respectively. Then the optimality
  of likelihood cross-validation implies that the 
  difference between $\log \widehat{BF}$ and $\log\widetilde{BF}$ is
  negligible, where  
$$
\widetilde{BF}=\frac{\pi_X(\alpha_r)L_X(\alpha_r)\pi_Y(\beta_s)L_Y(\beta_s)}{\pi(h_{r+s})L_0(h_{r+s})}.  
$$
If the priors and their parameters are chosen as discussed in Section
\ref{kp}, then, for example,
\beqn
\log
\pi_X(\alpha_r)&=&\log(2/\sqrt\pi)+\log(\hat\alpha/\alpha_r^2)-(\hat\alpha/\alpha_r)^2\\
&=&\log(2/\sqrt\pi)-\log(\alpha_r)-1+o_p(1)\\
&=&O_p(\log m),
\eeqn
with the last equality owing to the fact that the optimal bandwidth
$\alpha_r$ is of order $r^{-a}$ for $0<a<1$. Similar results are true
for the other two terms depending on priors. This entails that the
effect of the priors on $\log\widetilde{BF}$ is negligible
compared to the contribution from the likelihoods, as we will
see subsequently.

Defining $\hat
f_X\equiv \hat f(\,\cdot\,|\alpha_r,\X_T)$, $\hat f_Y\equiv \hat f(\,\cdot\,|\beta_s,\Y_T)$, $\hat f_{X,Y}\equiv \hat
f(\,\cdot\,|h_{r+s},\X_T,\Y_T)$ and 
$LR=L_X(\alpha_r)L_Y(\beta_s)/L_0(h_{r+s})$, we have 
\beq\label{firstlogLR}
\log(LR)=(m-r)\int\log\left(\frac{\hat f_X(x)}{\hat f_{X,Y}(x)}\right)dF_{m-r}(x)+(n-s)\int\log\left(\frac{\hat f_Y(y)}{\hat f_{X,Y}(y)}\right)dG_{n-s}(y),
\eeq
where $F_{m-r}$ and $G_{n-s}$ are the empirical cdfs of $\X_V$ and
$\Y_V$, respectively. Note that $\log(LR)$ depends intimately on entropy estimates of the form $-\sum_{i=1}^m\log\hat f(X_i)/m$. From (\ref{firstlogLR}) we have
\beq\label{logLR}
\log(LR)&=&(m-r)\left[KL(f,\hat f_{X,Y})-KL(f,\hat f_X)\right]+(n-s)\left[KL(f,\hat f_{X,Y})-KL(f,\hat f_Y)\right]\notag\\&&+\delta_1+\delta_2,
\eeq
where

$$
\delta_1=(m-r)\left\{\int\log\left(\frac{\hat f_X(x)}{\hat
    f_{X,Y}(x)}\right)\left[dF_{m-r}(x)-dF(x)\right]\right\}
$$
and 
$$
\delta_2=(n-s)\left\{\int\log\left(\frac{\hat f_Y(y)}{\hat
    f_{X,Y}(y)}\right)\left[dG_{n-s}(y)-dF(x)\right]\right\}.
$$

Under the conditions of \cite{Hall87}, the quantity
$\delta_1+\delta_2$ is negligible relative to the other terms in
$\log(LR)$, and 
\beq\label{expansion}
KL(f,\hat f_X)=C_fr^{-a}+o_p(r^{-a}),
\eeq
where $C_f$ is a positive constant depending on $f$ (and $K$) and
$a$ is a constant such that $0<a<4/5$. The other two Kullback-Leibler
discrepancies admit similar expansions, differing only with respect to
sample size. 
Expansion (\ref{expansion}) and the argument above imply that 
\beqn
\log(LR)&=&C_f\left\{(m-r)\left[\frac{1}{(r+s)^{a}}-\frac{1}{r^{a}}\right]+(n-s)\left[\frac{1}{(r+s)^{a}}-\frac{1}{s^{a}}\right]\right\}\\ &&+o_p\left(\frac{(m-r)}{r^{a}}+\frac{(n-s)}{s^{a}}\right). 
\eeqn
Obviously, each of the two terms in square brackets is negative, as 
desired in the present case where the two densities are the
same. Furthermore, if $r$ and $s$ tend to 
$\infty$ in such a way that $r/(r+s)$ converges to $q$ for $0<q<1$,
then 
\beq\label{nulllimit}\log(LR)\sim C_f\left\{\frac{(m-r)}{r^a}\cdot(q^a-1)+\frac{(n-s)}{s^a}\cdot((1-q)^a-1)\right\}.\eeq
Suppose that $r<mp$ and $s<np$ for $0<p<1$. Then since
$a<4/5$, the last quantity will tend to $-\infty$ as $m$, 
$n$, $r$ and $s$ tend to $\infty$. 

We turn now to the case where the alternative is
  true. To fix ideas, we define $f$ and $g$ to be different if and
  only if $\int|f-g|>0$. For $0<\lambda<1$, $f$ and $g$ different
  implies that $\int |f-(\lambda f+(1-\lambda)g|>0$ and $\int |g-(\lambda
  f+(1-\lambda)g|>0$, 
  inequalities that are necesssary for our consistency argument. Our
  new argument is essentially the same as in the null case until we
  arrive at (\ref{logLR}), which now becomes 
\beq\label{logLRalt}
\log(LR)&=&(m-r)\left[KL(f,\hat f_{X,Y})-KL(f,\hat
  f_X)\right]+(n-s)\left[KL(g,\hat f_{X,Y})-KL(g,\hat f_Y)\right]\notag\\
&&+\delta_1+\delta_2.
\eeq
The discrepancies $KL(f,\hat f_X)$ and $KL(g,\hat f_Y)$ tend to 0 in
probability under the conditions of \cite{Hall87}. If $r/(r+s)$ tends
to $q$, $0<q<1$, then $KL(f,\hat f_{X,Y})$ and $KL(g,\hat f_{XY})$ converge
in probability to $KL(f,qf+(1-q)g)$ and $KL(g,qf+(1-q)g)$,
respectively. Since each of $\int |f-(qf+(1-q)g)|$ and $\int
|f-(qf+(1-q)g)|$ is positive, it follows from Pinsker's inequality that
$KL(f,qf+(1-q)g)$ and $KL(g,qf+(1-q)g)$ are both positive. Therefore,
$\log(LR)$ is asympotic to $Am+Bn$ for positive constants $A$ and $B$,
and consistency is proven.

We close this section with the following remarks.
\bi

\item[R1.] Expression (\ref{logLRalt}), which is correct
    under both null 
  and alternative hypotheses, shows that the density estimates
  from the training data have the main  
  responsibility for getting the sign of $\log(LR)$ right.

\item[R2.] When the null is true, we wish $\log(LR)$ to be
  negative. This occurs  
  with high probability owing to 
  the fact that $KL(f,\hat f)$ tends to be smaller when the sample
  size on which $\hat f$ is based becomes larger. Therefore, to make it more
  likely that both $\left[KL(f,\hat f_{X,Y})-KL(f,\hat
  f_X)\right]$ and $\left[KL(f,\hat f_{X,Y})-KL(f,\hat f_Y)\right]$
  are negative, it stands to reason that $r/(r+s)$ should not be too
  close to either 0 or 1. 
\item[R3.] When the alternative is true, we want $\log(LR)$ to
  be positive. This is guaranteed if $\hat f_X$ is closer to $f$ (in
  the Kullback-Leibler sense) than is $\hat f_{X,Y}$, and $\hat f_Y$
  is closer to $g$ than is $\hat f_{X,Y}$. To ensure that this is
  true, one should make $r$ and $s$ as large as possible. 

\item[R4.] Assuming that the sign of $\log(LR)$ is correct, the weight of
  evidence in favor of the alternative is dictated by the sizes of
  $m-r$ and $n-s$. 

\item[R5.] When $n$ and $m$ are sufficiently large, a good way to
  ensure that both the sign and magnitude of $\log(LR)$ are suitable is to
  take $r\sim m/2$ and $s\sim n/2$.

\item[R6.] Expression (\ref{nulllimit}) entails that, under the null,
  $CVBF$ tends to 0 at a much faster rate than is typical for
  traditional Bayesian tests. Suppose, for example, that $r\sim m/2$
  and $s\sim n/2$. Then $\log(LR)\sim -(C_1 m^{1-a}+C_2n^{1-a})$ for
  positive constants $C_1$ and $C_2$ and $0<a<4/5$. In contrast, when
  one uses a nonparametric Bayesian procedure in which the null model
  is nested 
  within the alternative, the log-Bayes factor typically diverges to
  $-\infty$ at a rate that is only logarithmic in the sample size; see, for
  example \cite{MRM}.  

\ei

\section{Choice of training set size}\label{tv}

We now discuss the choice of training set sizes $r$ and $s$ for given
sets of data. Two competing ideas are at play when choosing these
quantities. First of all, it is desired that the KDEs from the
training data be good representations of $f$ and $g$, a desire that
calls for large $r$ and $s$. On the other hand, we would like as much
data as possible for computing the Bayes factor, which asks that $m-r$
and $n-s$ be large. To balance these two considerations it seems
intuitively reasonable that one choose $r=[m/2]$ and $s=[n/2]$,
where $[x]$ denotes the integer closest  to but not larger than
$x$. Indeed, when $m+n$ is smaller than 5000 or so, these are our
suggested default choices of $r$ and $s$.  The main
reason that $r=[m/2]$ and $s=[n/2]$ are not suggested for very large
sample sizes is that they {\it maximize} the length of time needed to
compute the Bayes factor. To explain why, consider the marginal based on just $X_1,\ld,X_m$. 
This marginal requires calculation of $m-r$ kernel estimates, each of which 
involves $r$ additions. So, a total of 
$r(m-r)$ operations are required for each likelihood evaluation, and
this number is maximized when $r=m/2$. This result is of particular
interest for extremely large datasets. In this case it is 
unlikely that half of the full dataset is required for
computing a good training density, and hence significant reductions in
computing time can be gained by choosing a training set size that is
much smaller than the validation set size.  

Aside from computing issues, there is another reason why $r=[m/2]$ and
$s=[n/2]$ are not suggested for very 
large data sets. Doing so is undoubtedly not 
optimal from the point of view of producing good Bayes
factors. Expressions (\ref{nulllimit}) and (\ref{logLRalt}) suggest that
the magnitude of log-$CVBF$ tends to decrease with an increase in the
training set size, this being true under both null and alternative
hypotheses. Arguing on a more intutive level, suppose that
$m=n=50,000$. Choosing 
$r=s=25,000$ is undoubtedly overkill for obtaining reliable KDEs and
reduces the number of data 
available for computing the Bayes factor(s). A whole range of much
smaller choices of $r=s$ will produce high quality KDEs and leave more
data for computing CVBFs. 

Dependence of Bayes procedures on tuning parameters or
hyperparameters, especially in nonparametric settings, is not at all
unusual. For example, the P\'olya tree method of \cite{holmes2015two}
relies upon specifying the prior precision parameter $c$. The authors of
that article state that values of $c$ between $1$ and $10$ work well
in practice, but they also recommend checking the sensitivity of their
Bayes factor to choice of $c$. When employing our CVBF methodology the
training set sizes may be regarded as tuning parameters, and as with
any Bayes procedure it is recommended that one investigate sensitivity
of $CVBF$ to different choices for $(r,s)$.  If all the
Bayes factors computed are in basic agreement, then the decision is
clear.  

To deal with cases where Bayes factors corresponding to different
choices of $(r,s)$ are not in agreement, we propose that one treat
$(r,s)$ as a parameter that has a prior distribution. To simplify
matters, we assume that $m=n$ and $r=s$. Now, divide each
data set into training and validation sets of size $K$ and $m-K$,
respectively. For sample sizes that are not extremely large one could
take $K=[m/2]$, and otherwise $K$ could be some reasonable upper bound
on $r$. Let ${\cal T}$ be some subset of training set sizes between 1
and $K$, and suppose that prior probabilities $p(r)$ are assigned to
the elements of ${\cal T}$ in such a way that $\sum_{r\in{\cal
    T}}p(r)=1$. (A good default choice would be to assume that the
training set sizes are equally likely.) For each
$r\in{\cal T}$ we randomly select $r$ 
values (without replacement) from each of the two sets of training
data.  From these two data sets of size $r$ we determine kernel models
$M_{X,r}$, $M_{Y,r}$ and $M_{0r}$, as described in Section
\ref{method}. From these we may compute marginal likelihoods $m_{0,r}$ and
$m_r$ corresponding to the null and alternative hypotheses,
respectively, from the validation data, again as described in Section 
\ref{method}. Assuming that the prior probability of $H_0$ is the same
for each training set size, a Bayes
factor for comparing the alternative and null models is 
\begin{equation}\label{AverOverTrains}
\frac{\sum_{r\in{\cal T}}p(r)m_r}{\sum_{r\in{\cal T}}p(r)m_{0,r}}.
\end{equation}
Importantly, this methodology is in strict adherence to Bayesian
principles, inasmuch as all models are formulated from training data,
and the two hypotheses are assessed from the same set of validation
data, which is independent of the training data. This method will be
illustrated in our real data example in Section \ref{data}.

\section{Simulations}\label{sims}

Initially we provide evidence that our cross-validation method of
selecting a bandwidth is efficient in the sense of \cite{Hall87}.
Suppose we have independent random samples $\X_T=(X_1,\ld,X_r)$
  and $\X_V=X_{r+1},\ld,X_m$, each from the same density $f$. We
  wish to select the bandwidth of the KDE $\hat
  f(\,\cdot\,|h,\X_T)$. We do so in two ways, by determining the maximizers of the criteria defined in (\ref{loo}) and (\ref{ourversion}). Table \ref{bandwidths} provides
results for a setting in which data are 
drawn from $f\equiv\hbox{standard Cauchy density}$ and the kernel used
is the Hall 
kernel, as defined in (\ref{Hallkern}). One hundred replications of each
of two cases were performed: $(r,m-r)=(200,300)$ and
$(r,m-r)=(400,600)$. For each data set three bandwidths were determined:
the maximizers of (\ref{loo}) and (\ref{ourversion}) and the minimizer of the 
Kullback-Leibler discrepancy between KDEs and the true Cauchy density.
 
\begin{table}
\begin{center}
\begin{tabular}{|r|c|c|c|c|c|c|}
\hline
$(r,m-r)$ &Mean $\hat h_{LO}$& SD $\hat h_{LO}$&Mean $\hat h_{CV}$&SD $\hat h_{CV}$&Mean $\hat h_{KL}$& SD
$\hat h_{KL}$\\
\hline
$(200,300)$ &0.320 &0.080 & 0.314 &0.073 &0.313 & 0.025   \\
\hline
$(400,600)$ &0.277 &0.066    & 0.265 &0.044& 0.268& 0.020 \\
\hline
\end{tabular}
\caption{ {\it Means and standard deviations of cross-validation
    bandwidths and Kullback-Leibler optimal bandwidths. The bandwidths $\hat h_{LO}$ and $\hat h_{CV}$ maximize (\ref{loo}) and (\ref{ourversion}), respectively, and $\hat h_{KL}$ optimizes Kullback-Leibler loss. Results are based on 100 replications for each choice
of $(r,m-r)$.}}
\label{bandwidths} 
\end{center}
\end{table}

\begin{figure}[t]

\begin{center}

{\includegraphics[height=3in]{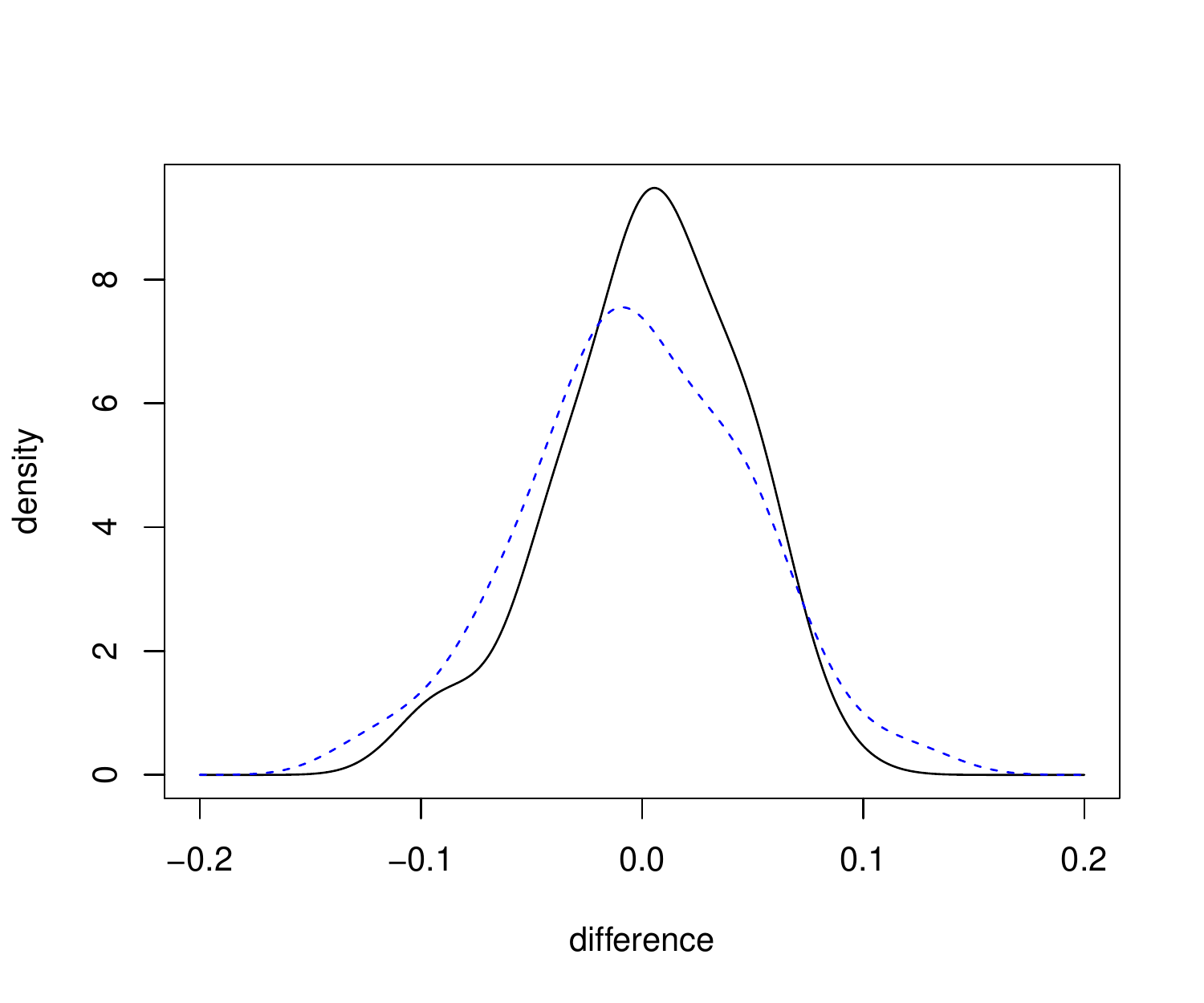}}

\caption{\it Kernel density estimates of bandwidth differences distributions. The
  solid line is a KDE of the difference between the Kullback-Leibler optimal bandwidth and the maximizer of (\ref{ourversion}), while the dashed blue line is a KDE of the difference between the Kullback-Leibler optimal bandwidth and the maximizer of (\ref{loo}). Results are based on 100 replications and the case where $r$ is 400 and $m-r$ is 600.} 

\label{densbw}

\end{center}

\end{figure}

The cross-validation bandwidths behave in accordance with the theory described in Section \ref{cons}. At a given
$(r,m-r)$, the means of $\hat h_{LO}$ and 
$\hat h_{CV}$  are approximately equal to the mean of the optimal
bandwidths. Furthermore, the standard deviations of the 
cross-validation bandwidths decrease when $r$ amd $m-r$
increase. Figure \ref{densbw} illustrates that the bandwidth maximizing (\ref{ourversion}) tends to be closer to the Kullback-Leibler optimal bandwidth than is the maximizer of leave-one-out cross-validation.  

As noted previously, \cite{Hall87} proved that when
  the tails of the underlying density are sufficiently heavy and one
  uses a Gaussian kernel, then likelihood cross-validation chooses a
  bandwidth that diverges to infinity as sample size increases without
  bound. To illustrate this point, we repeated the previous simulation
  using a Gaussian kernel instead of (\ref{Hallkern}). For each data
  set, $L(h)$ was maximized over the interval $(0.001,30)$. At
  $(r,m-r)=(200,300)$, the average value of the cross-validation 
  bandwidths was 13.4, and 16 of the 100 values were 30. At
  $(r,m-r)=(400,600)$ the average bandwidth was 12.8 and 6 of the 100
  were 30. These bandwidths are obviously much too big to provide
  reasonable estimates of the true density.

We turn now to simulations investigating various aspects of our CVBF
methodology. Part of this investigation addresses how our test fares
in comparison to the Kolmogorov-Smirnov (KS) test and to the P\'olya
tree test of \cite{holmes2015two}. For each case where the P\'olya tree test is run, the precision
parameter $c$ is taken to be 1. We do this for two
reasons. Primarily, this choice proved to be successful in the study
of \cite{holmes2015two}. Secondly, choosing $c$ to be closer to 0 seems to
have the effect of making the test depend less on the centering
distribution utilized and more on the empirical cdf, which is what
would be desired in the non-parametric setting \citep{hanson2006inference}. 

For the null case, we
generate data from a standard normal  
distribution, taking $m=n$ for sample sizes 200, 400 and 800. For the CVBF
training set sizes we took $r=s$, with $r=50$, $75$ and $112$ at
sample sizes $200$, $400$, and $800$, respectively. 
So, the training 
set size increases by fifty percent when the sample size doubles. The
value of CVBF for a given 
replication was the geometric mean of CVBFs corresponding to 30 pairs of
randomly selected training sets, and 1500 replications were
performed at each $n$. The kernel 
$K_0$ was used for all our simulations, and the prior for each
bandwidth was (\ref{prior}) with $\gamma$ equal to the maximizer of the corresponding likelihood.

\begin{figure}[!htbp]
\begin{center}
{\includegraphics[height=3in]{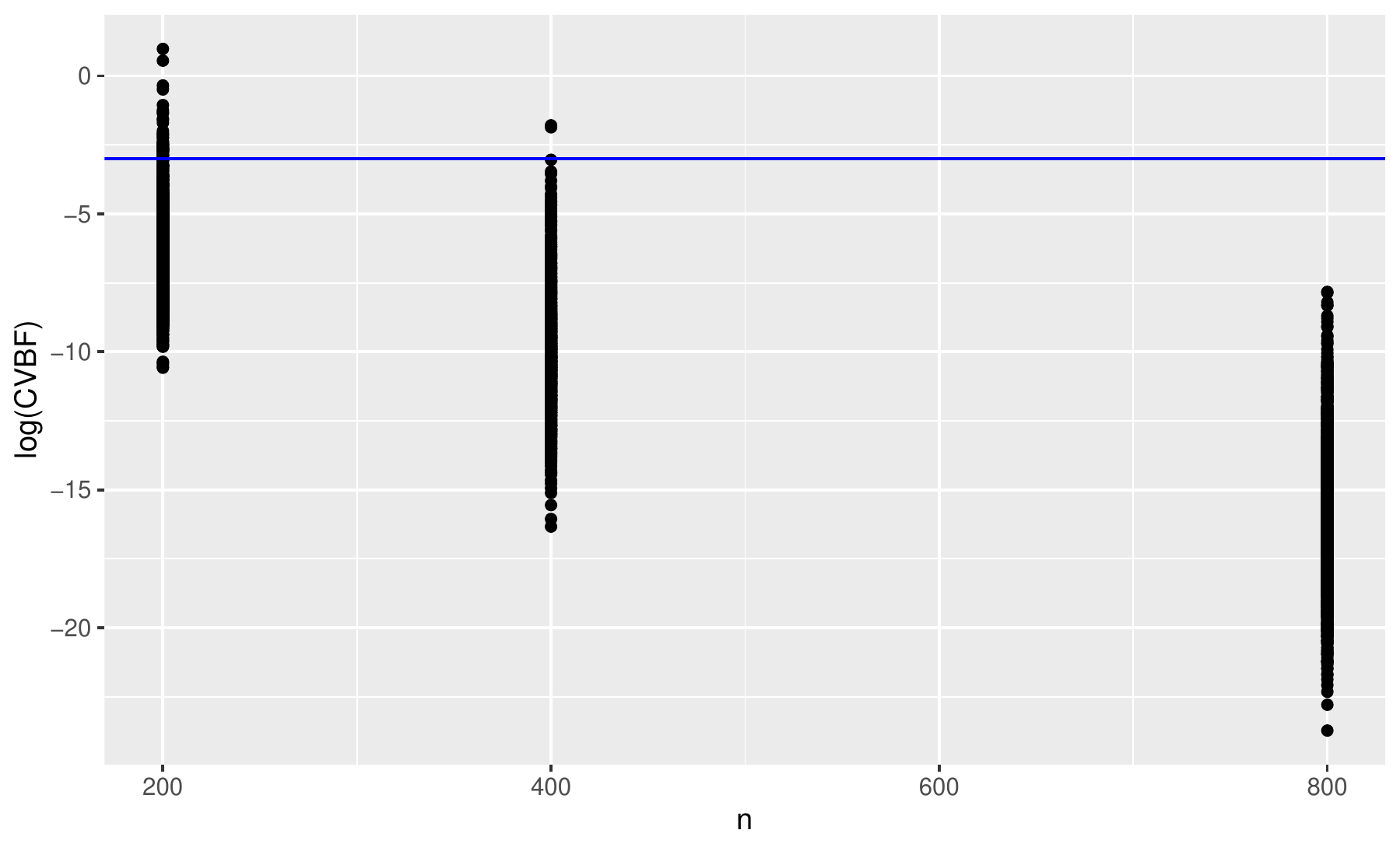}}
\caption{\it Values of $\log(CVBF)$ when the null hypothesis is true.
  Each point corresponds to $X$ and $Y$ samples each of size $n$ from a
  $N(0,1)$ distribution. The standard deviations of the log-Bayes factors are $1.60$, $1.95$ and $2.41$, for $n=200$, $400$, and $800$, respectively.}
\label{null}
\end{center}
\end{figure}

\begin{figure}[!htbp]
\begin{center}
{\includegraphics[height=3in]{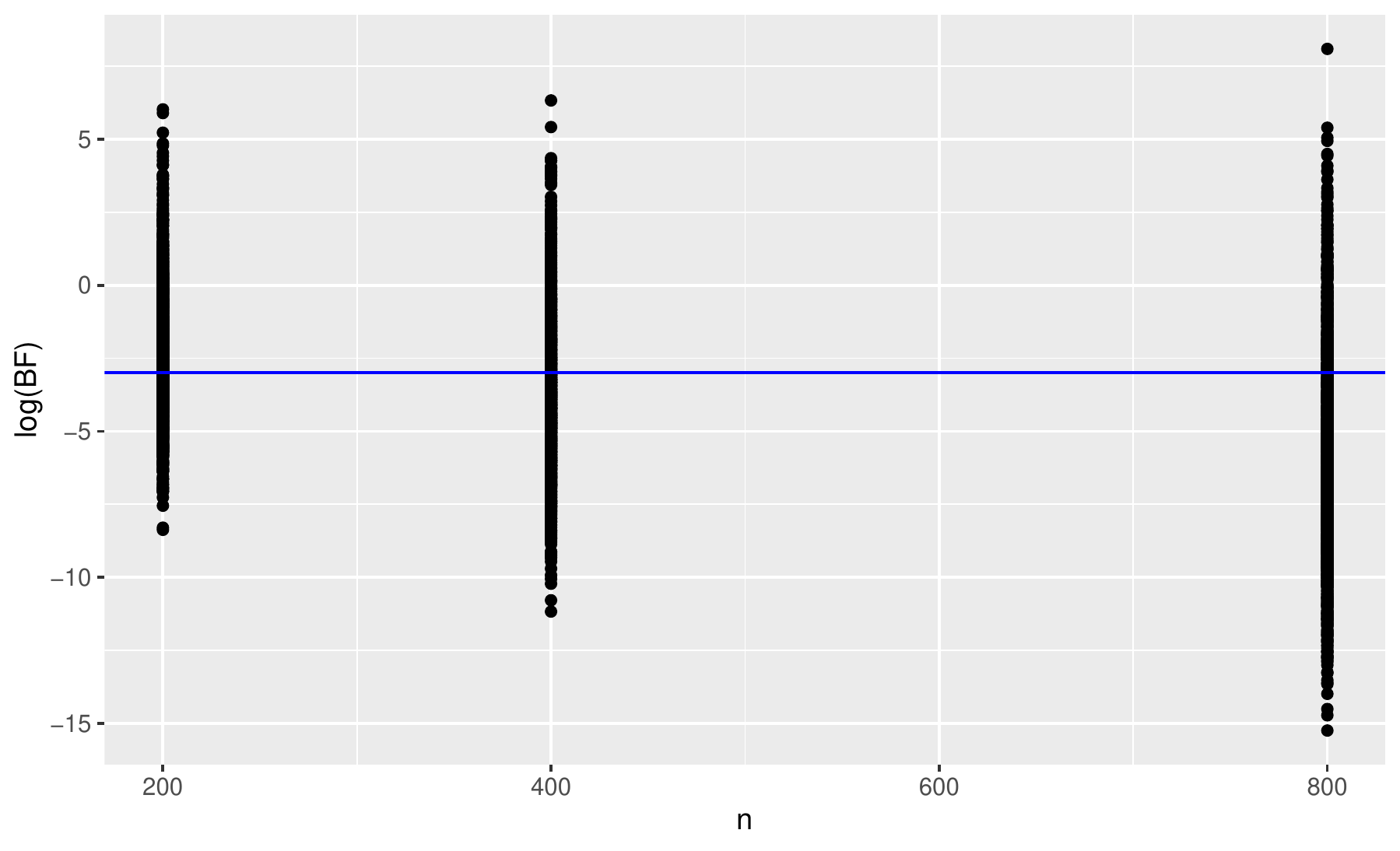}}
\caption{\it Values of $\log(BF)$ for the P\'olya tree when the null hypothesis is true.
  Each point corresponds to $X$ and $Y$ samples each of size $n$ from a
  $N(0,1)$ distribution. The standard deviations of the log-Bayes factors are $2.05$, $2.52$ and $3.31$  for $n=200$, $400$, and $800$, respectively.}
\label{null2}
\end{center}
\end{figure}

The results are shown in Figures \ref{null} and \ref{null2}.  All but two of the 4500 values of $\log(CVBF)$ computed were smaller than 0. At $n=400$, all 1500 replications 
produced a value of $\log(CVBF)$ that was smaller than 0, and just two values larger than $-\log(20)$, a value considered to be the threshold for 
``strong'' evidence in favor of the null hypothesis
\citep{KassRaft}.  At $n=800$ all 1500 values of $\log(CVBF)$ were smaller than $-\log(20)$, and at $n=200$ all but 46 values of $\log(CVBF)$ were below this threshold. The near linear decrease in the estimates of 
$E(\log(CVBF))$ is 
evidence for the exponential rate of convergence of $CVBF$ that was
discussed in Remark R6. The P\'olya tree Bayes factors do not behave as well as the CVBFs. For example, at $n=400$, the median $\log(CVBF)$ is $-10.26$, while the median P\'olya tree $\log(BF)$ is but $-4.06$. At $n=800$, $5$\% of the P\'olya tree $\log(BF)$ values are actually larger than 0, while the {\it largest} $\log(CVBF)$ is $-7.83$. When a particular model is true, we desire that a Bayes factor provide the strongest possible evidence in favor of that model, and on this score CVBF has outperformed the P\'olya tree method in this example. 

Under the alternative hypothesis, we use a version of BayesSim, as
proposed by \cite{BSim}, for data generation. Here, the $X$ sample is drawn from a density $f$.
To obtain the $Y$ sample, we first draw $p$ from
beta$(1/2,1/2)$, a beta distribution with both
parameters equal to 1/2, and then the $Y$ sample is drawn from a
mixture of the form $(1-p)f(x)+pg(x)$, where $g$ is different from $f$. 
This approach allows one to infer the
behavior of CVBF for mixing proportions $p$ ranging from 0 to
1, where the discrepancy between the $X$ and $Y$ densities increases with $p$. Sample sizes of $m=n=280$ were considered, the training set sizes 
were selected to be 120, and 500 values of $p$ were selected for each choice of $(f,g)$. 
For a given replication, thirty random splits for each of $X$ and $Y$ were used. For each pair of data sets the data were centered and scaled before applying the P\'olya tree test.  The sample median of the  combination of the two data sets was subtracted from every value and then this difference was divided by $\text{IQR}/1.35$, where IQR is the interquartile range of the combined data.

The simulations just described were conducted in four different settings in each of which $f$ and $g$ differ in a particular way:
\bi \item[] {\it Scale change:} The densities $f$ and $g$ are $\phi$ (standard normal) and $\phi(x/2)/2$, respectively, and hence differ with respect to scale. 
\item[] {\it Location shift:} The densities $f$ and $g$ are standard
Cauchy, $f_C$, and $f_C(x+1)$, respectively, and so differ with respect to location.  

\item[] {\it Distributions with different tail behavior:} Here $f$ and $g$ are 
$f_C$ and $0.6745\phi(0.6745x)$, respectively. Given $p$, the mixture
density in this case has the same median and interquartile range as the 
standard Cauchy, and so the densities of the $X$ and $Y$ samples are different but have the
same location and scale. 
\item[] {\it Different distributions with same finite support:} The densities $f$ and $g$ are 
  $U(0,1)$ (uniform on the interval $(0,1)$) and beta$(1/2,1/2)$, respectively.   
\ei

\begin{figure}
    \centering
    \subfloat[P\'olya tree Bayes factors when a standard normal is used for quantiles. ]{{\includegraphics[width=6cm]{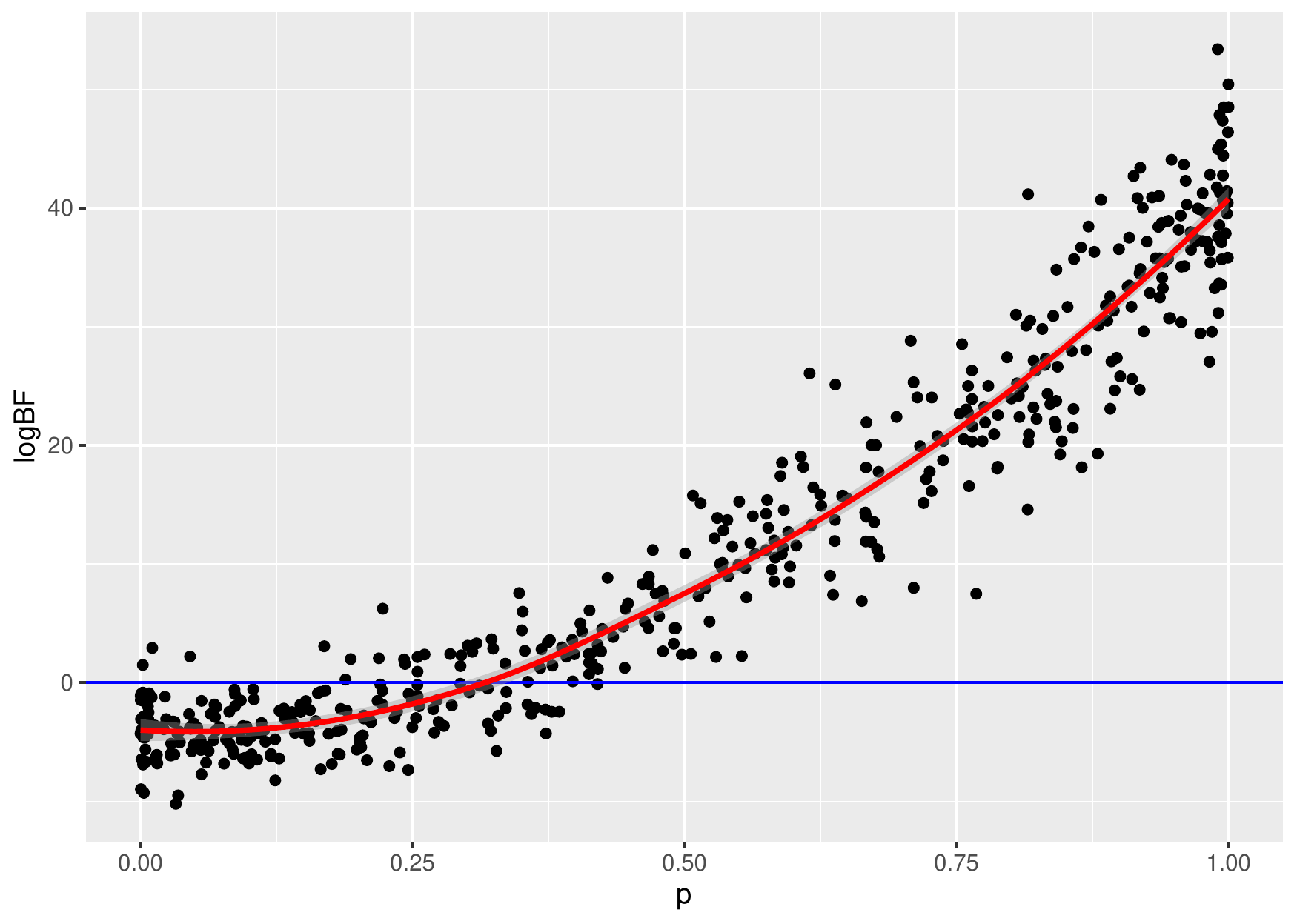} }}
    \qquad
    \subfloat[P\'olya tree Bayes factors when a standard Cauchy is used for quantiles. ]{{\includegraphics[width=6cm]{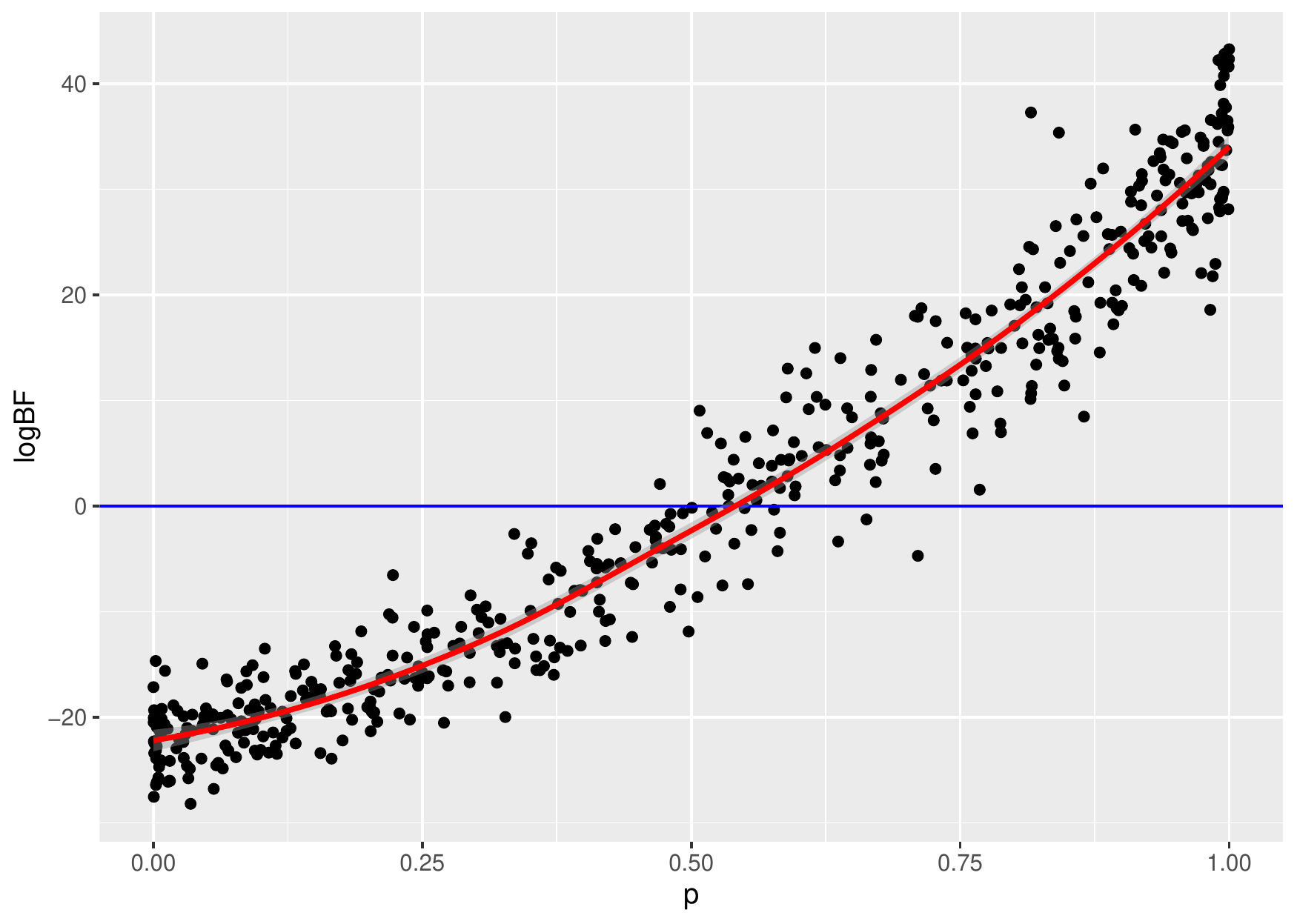} }}
    \qquad
    \subfloat[KS test log $P$-values]{{\includegraphics[width=6cm]{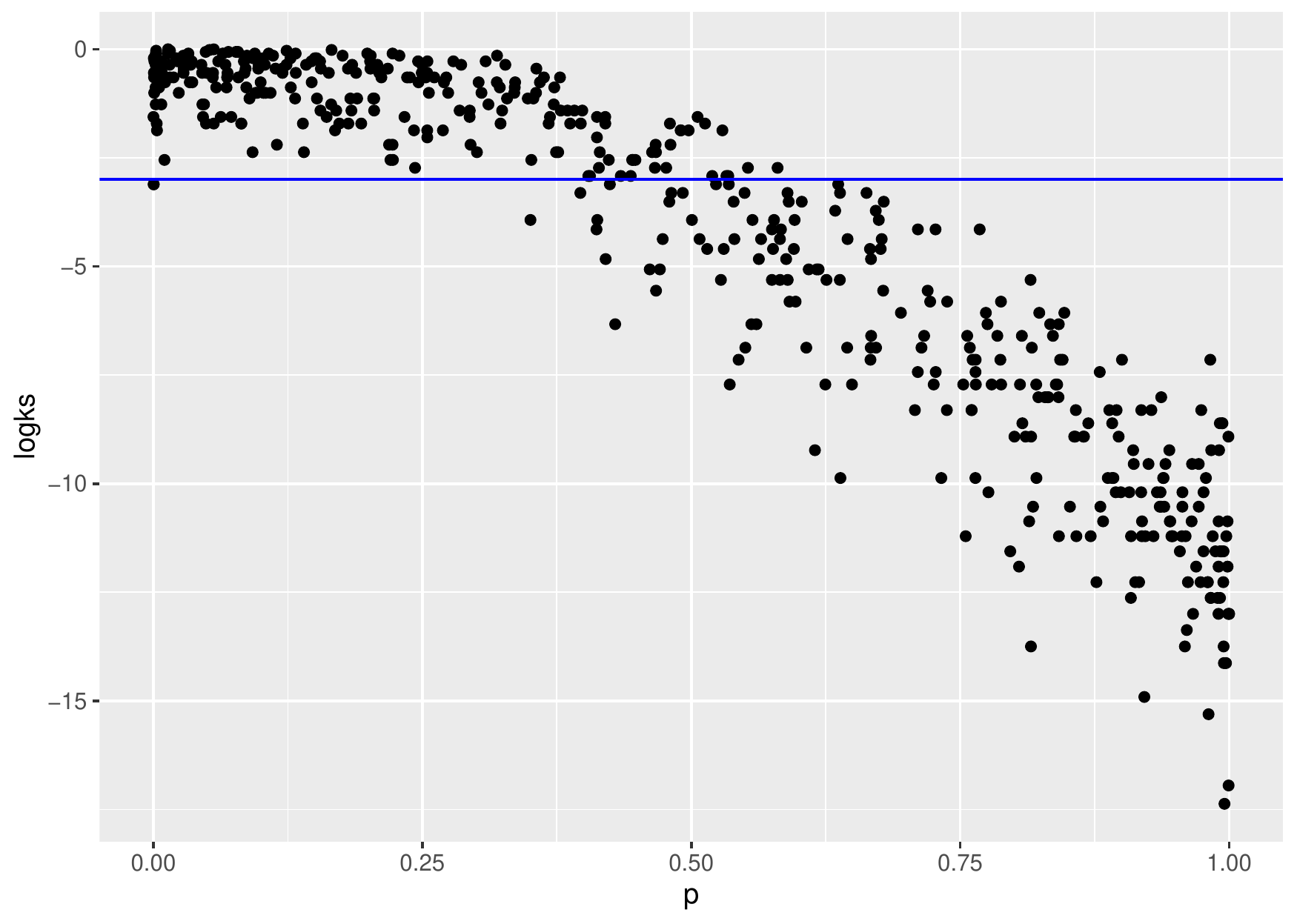} }}
    \qquad
    \subfloat[Cross-validation Bayes factors. ]{{\includegraphics[width=6cm]{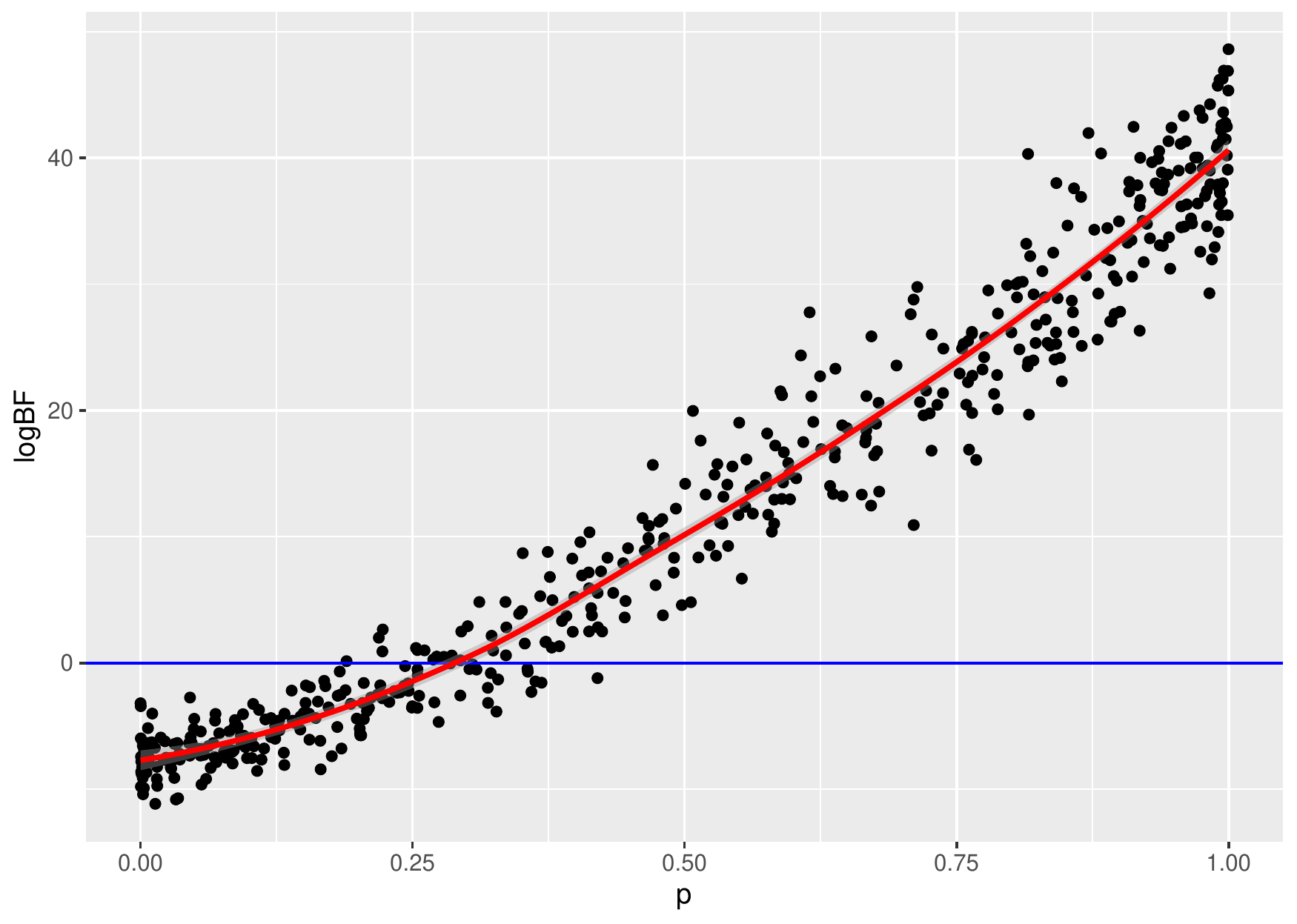} }}
    \caption{Comparison of tests in the scale change case. The densities $f$ and $g$ are both normal in this case.}
    \label{fig:ScaleShiftComp}
\end{figure}

\begin{figure}
    \centering
    \subfloat[P\'olya tree Bayes factors when a standard normal is used for quantiles. ]{{\includegraphics[width=6cm]{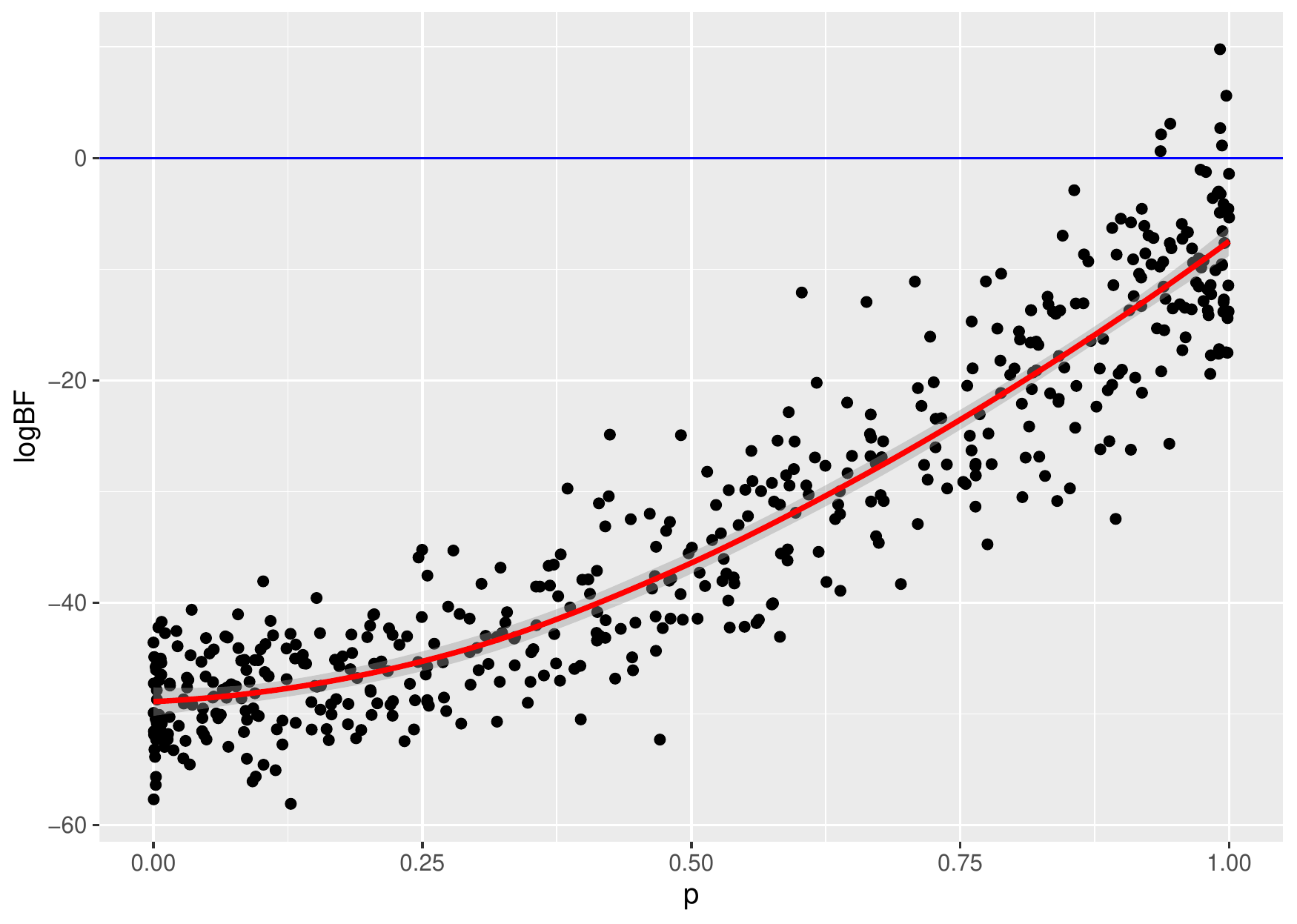} }}
    \qquad
    \subfloat[P\'olya tree Bayes factors when a standard Cauchy is used for quantiles. ]{{\includegraphics[width=6cm]{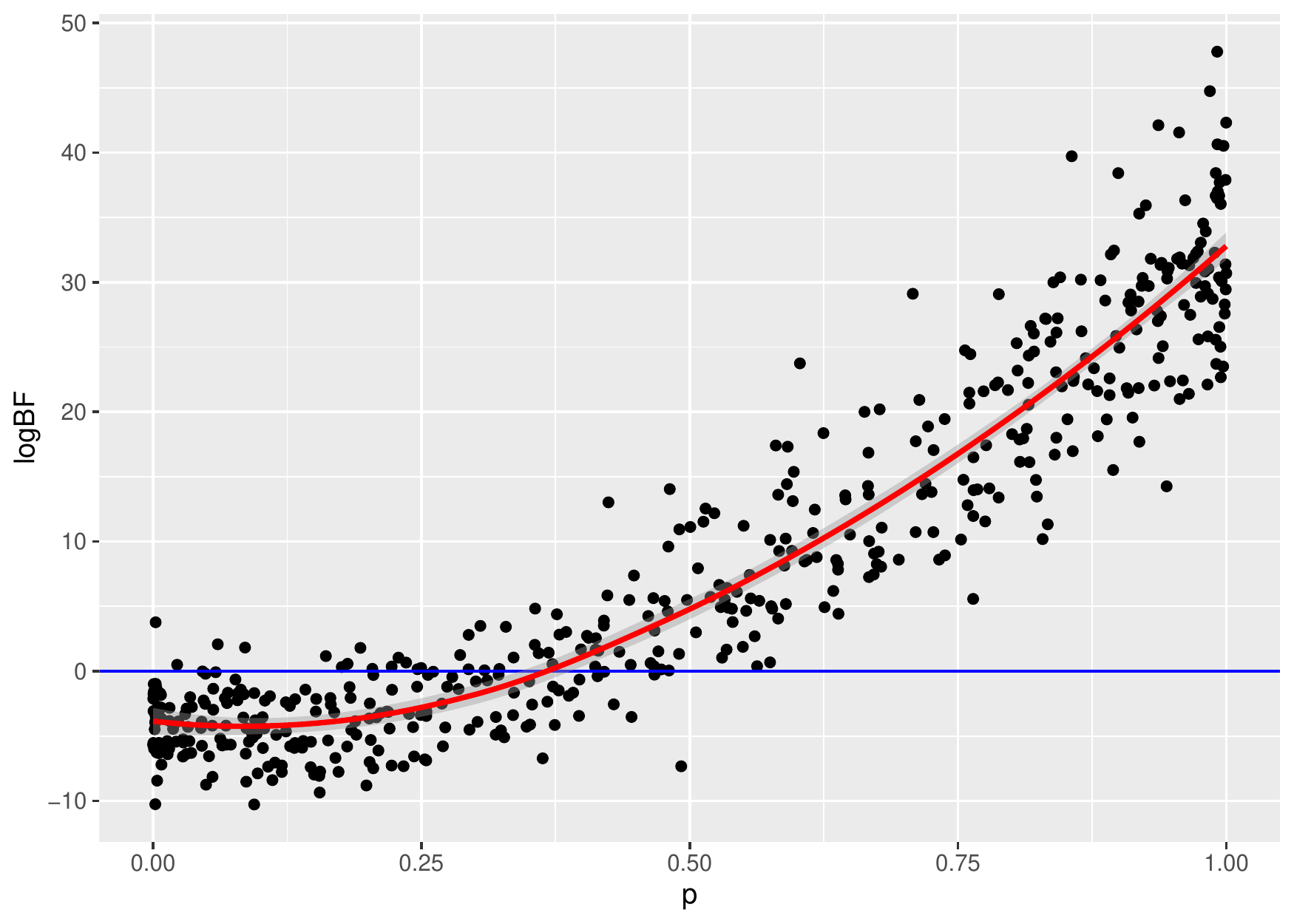} }}
    \qquad
    \subfloat[KS test log $P$-values]{{\includegraphics[width=6cm]{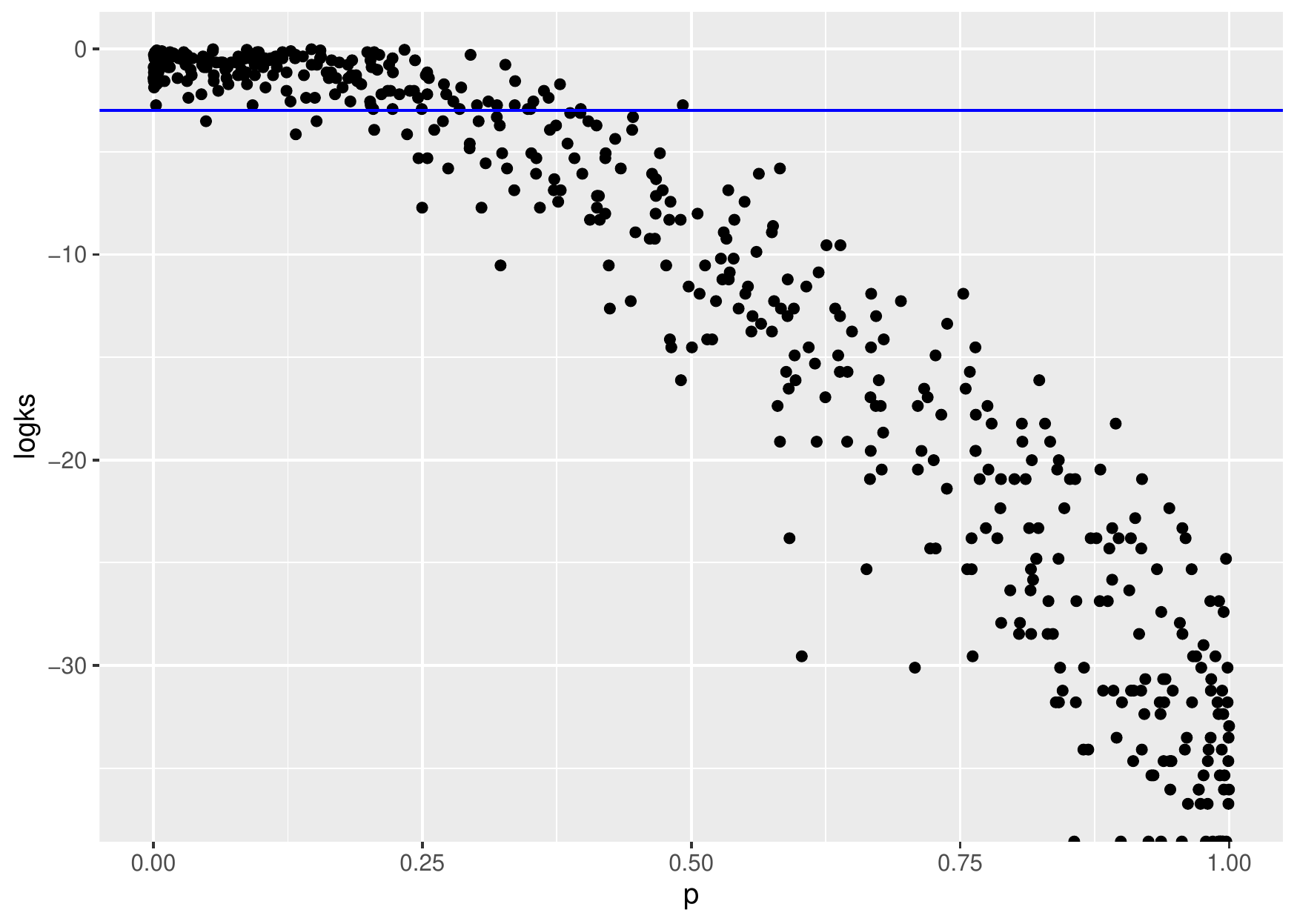} }}
    \qquad
    \subfloat[Cross-validation Bayes factors. ]{{\includegraphics[width=6cm]{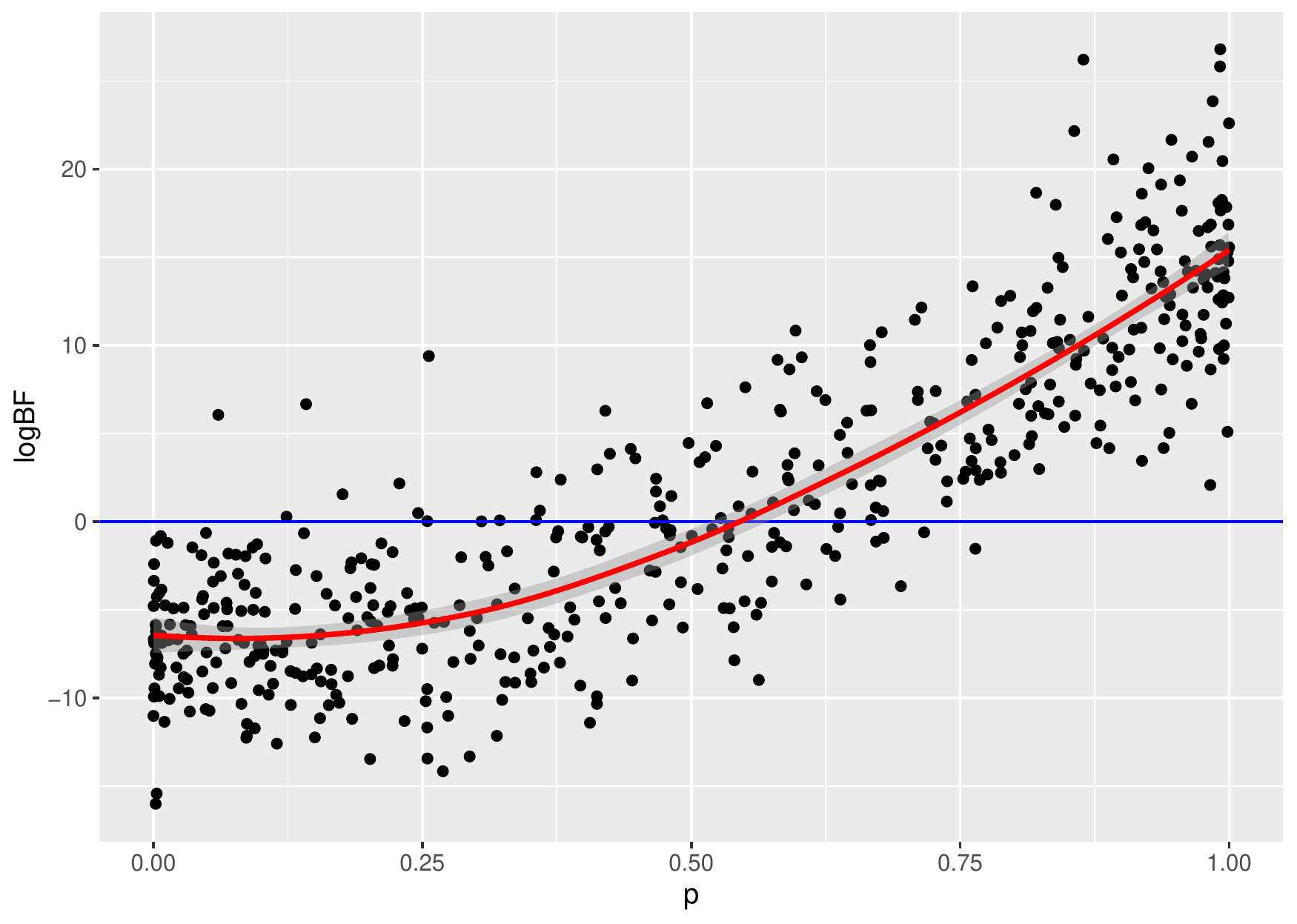} }}
    \caption{Comparison of tests in the location shift case.  The densities $f$ and $g$ are both Cauchy in this case.}
    \label{fig:LocationShiftComp}
\end{figure}

\begin{figure}
    \centering
    \subfloat[P\'olya tree Bayes factors when a standard normal is used for quantiles. ]{{\includegraphics[width=6cm]{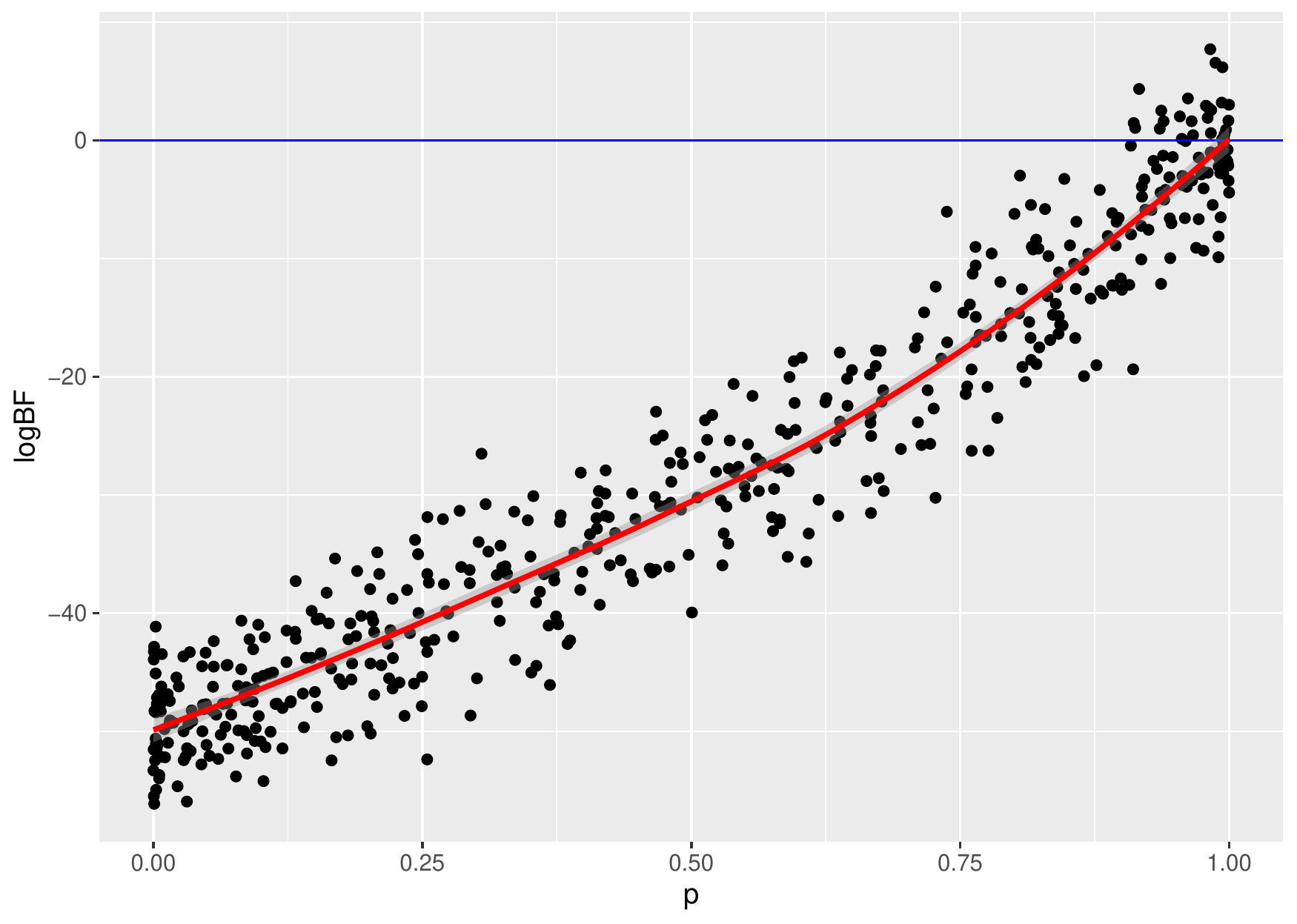} }}
    \qquad
    \subfloat[P\'olya tree Bayes factors when a standard Cauchy is used for quantiles.  ]{{\includegraphics[width=6cm]{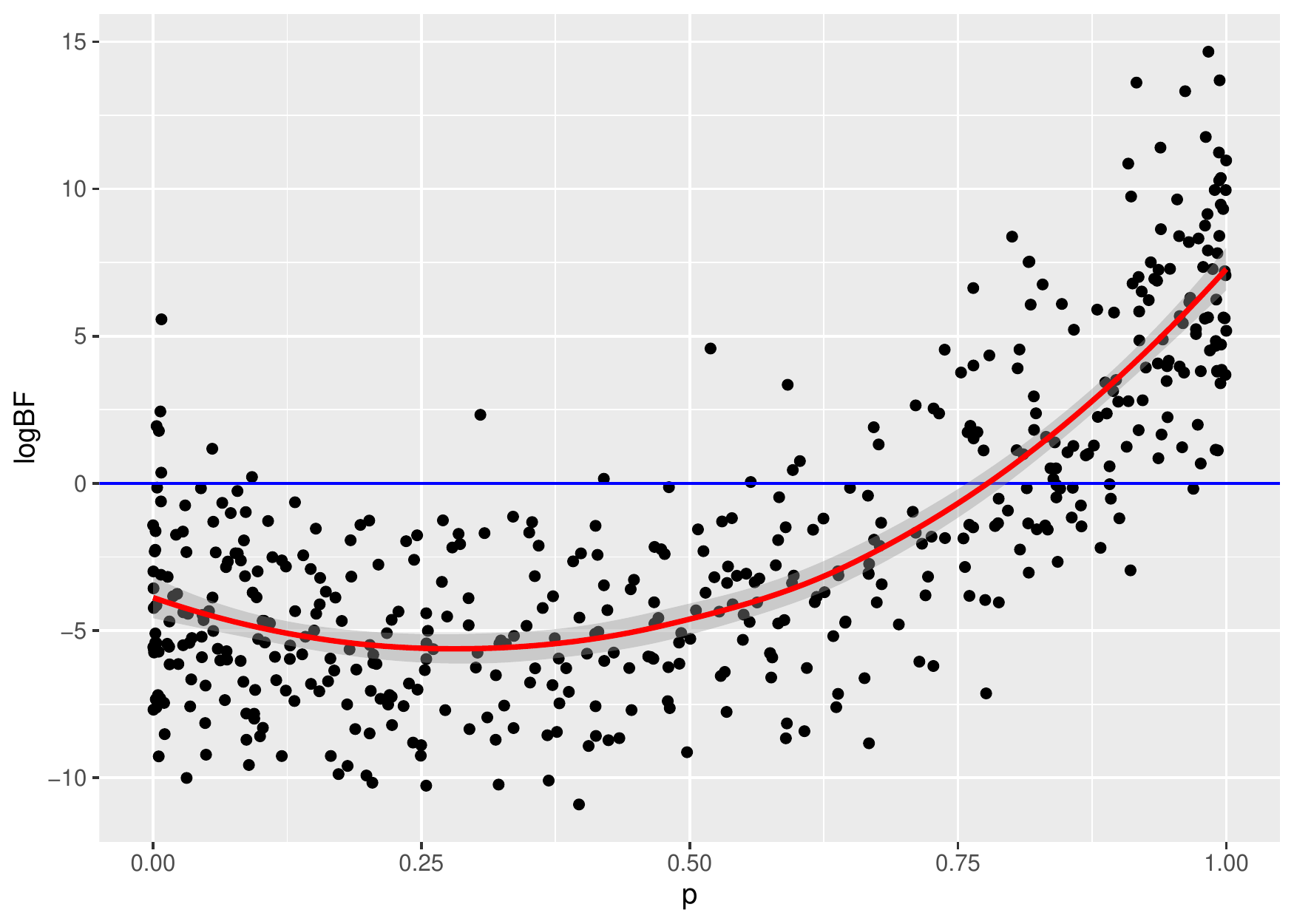} }}
    \qquad
    \subfloat[KS test log $P$-values]{{\includegraphics[width=6cm]{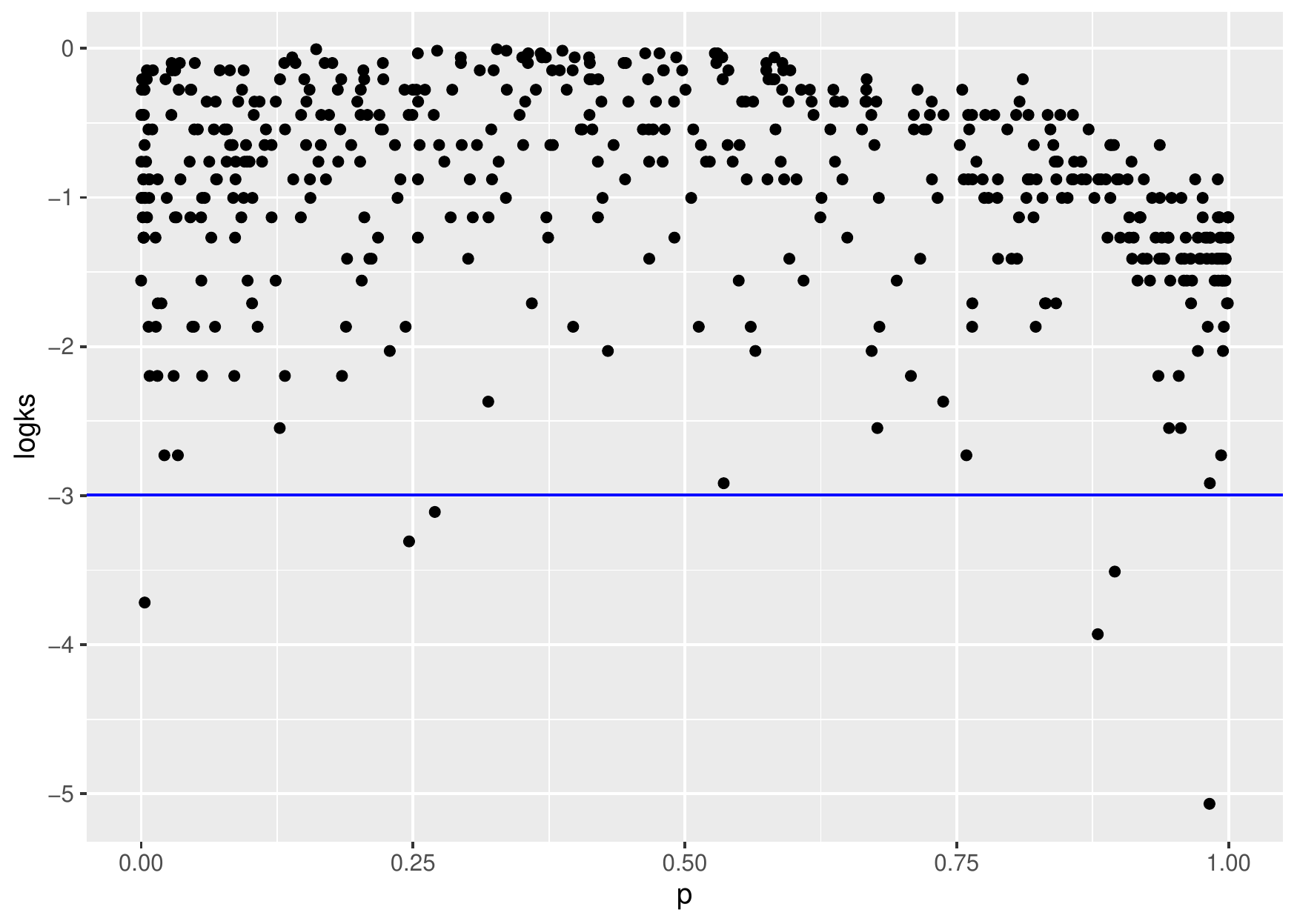} }}
    \qquad
    \subfloat[Cross-validation Bayes factors. ] {{\includegraphics[width=6cm]{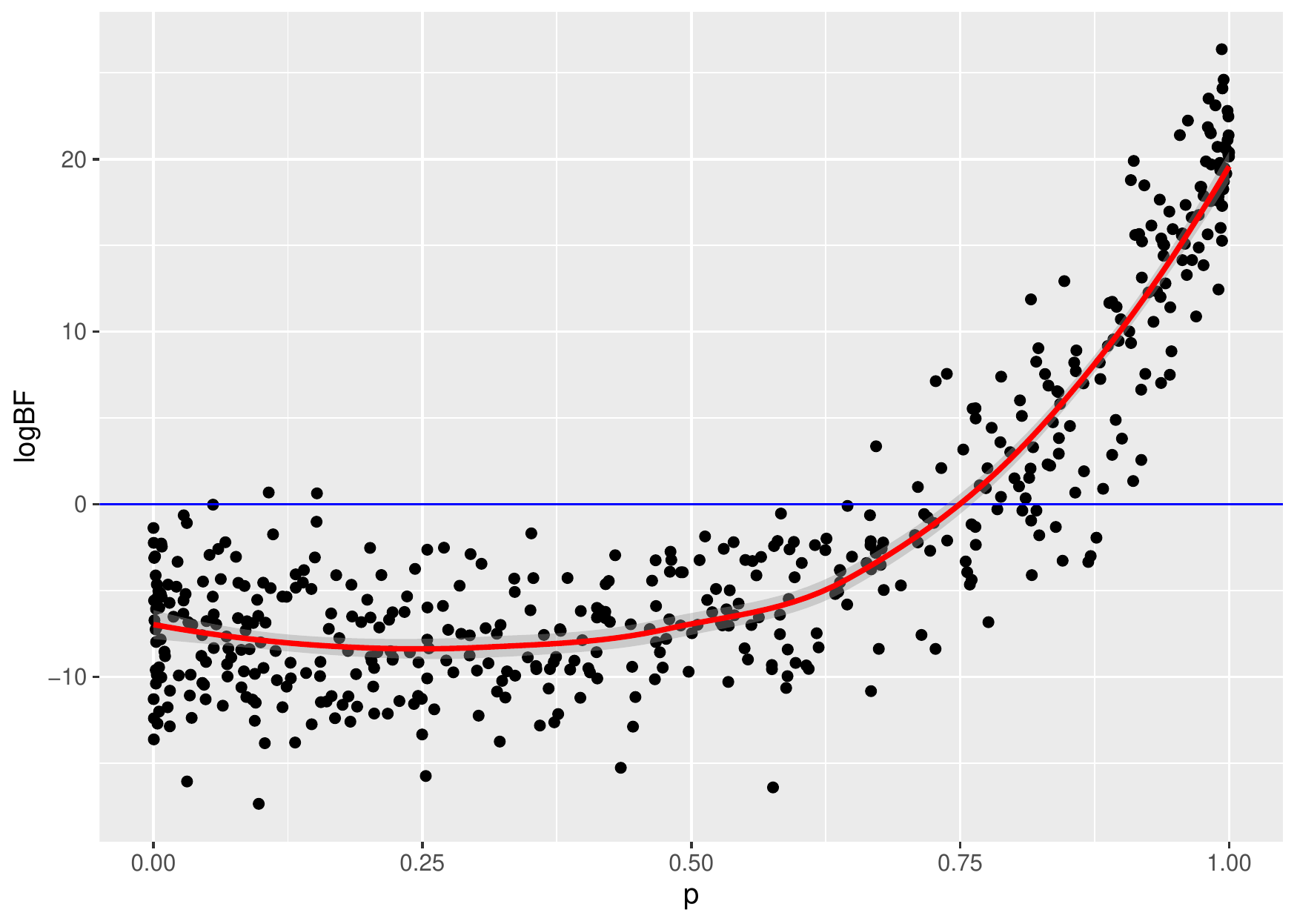} }}
    \caption{Comparison of tests in the case of distributions with different tail behavior. Here $f$ and $g$ are Cauchy and normal, respectively.}
    \label{fig:TailDiffComp}
\end{figure}

\begin{figure}
    \centering
    \subfloat[P\'olya tree Bayes factors when a standard normal is used for quantiles.]{{\includegraphics[width=6cm]{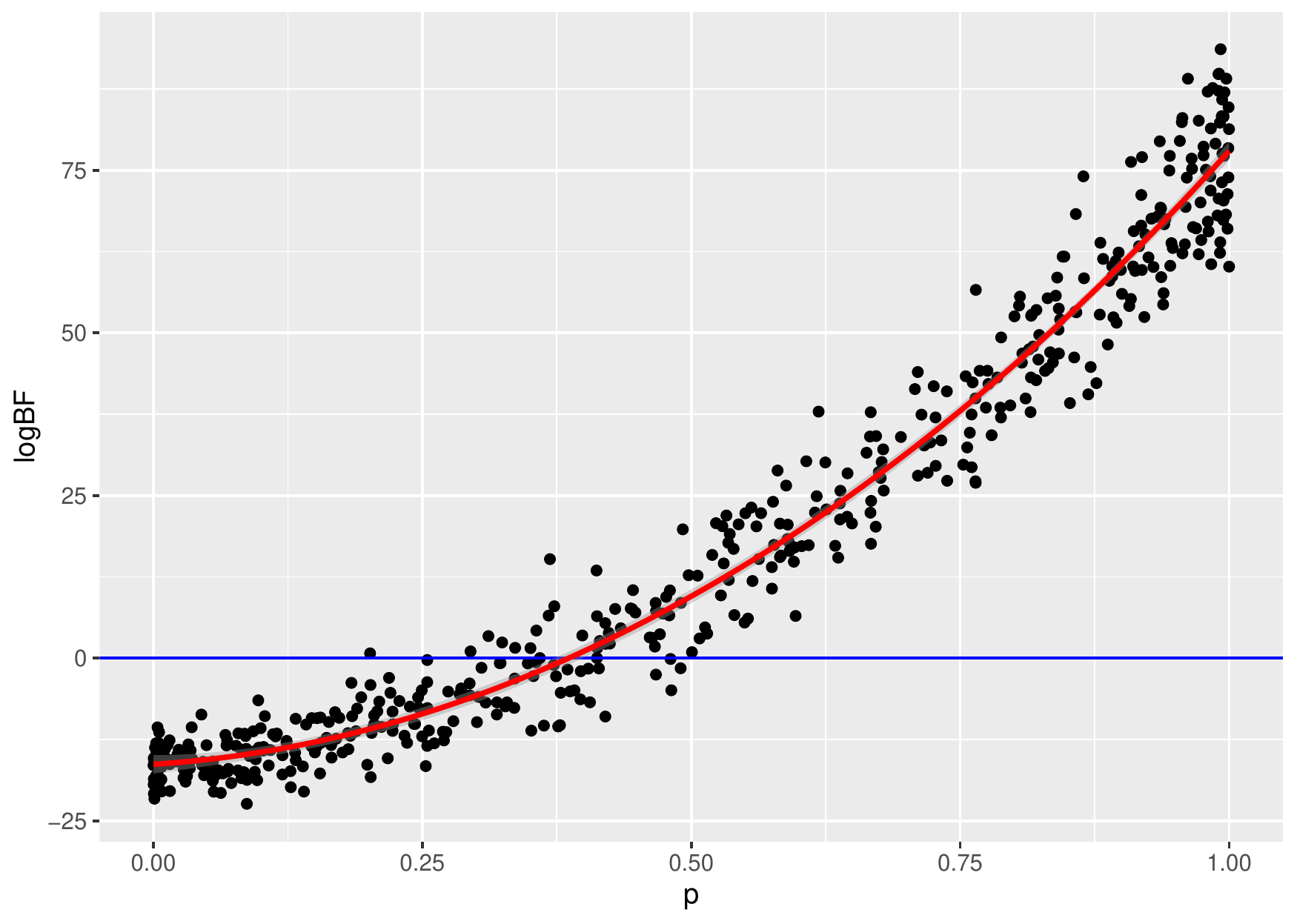} }}
    \qquad
    \subfloat[P\'olya tree Bayes factors when a standard Cauchy is used for quantiles. ]{{\includegraphics[width=6cm]{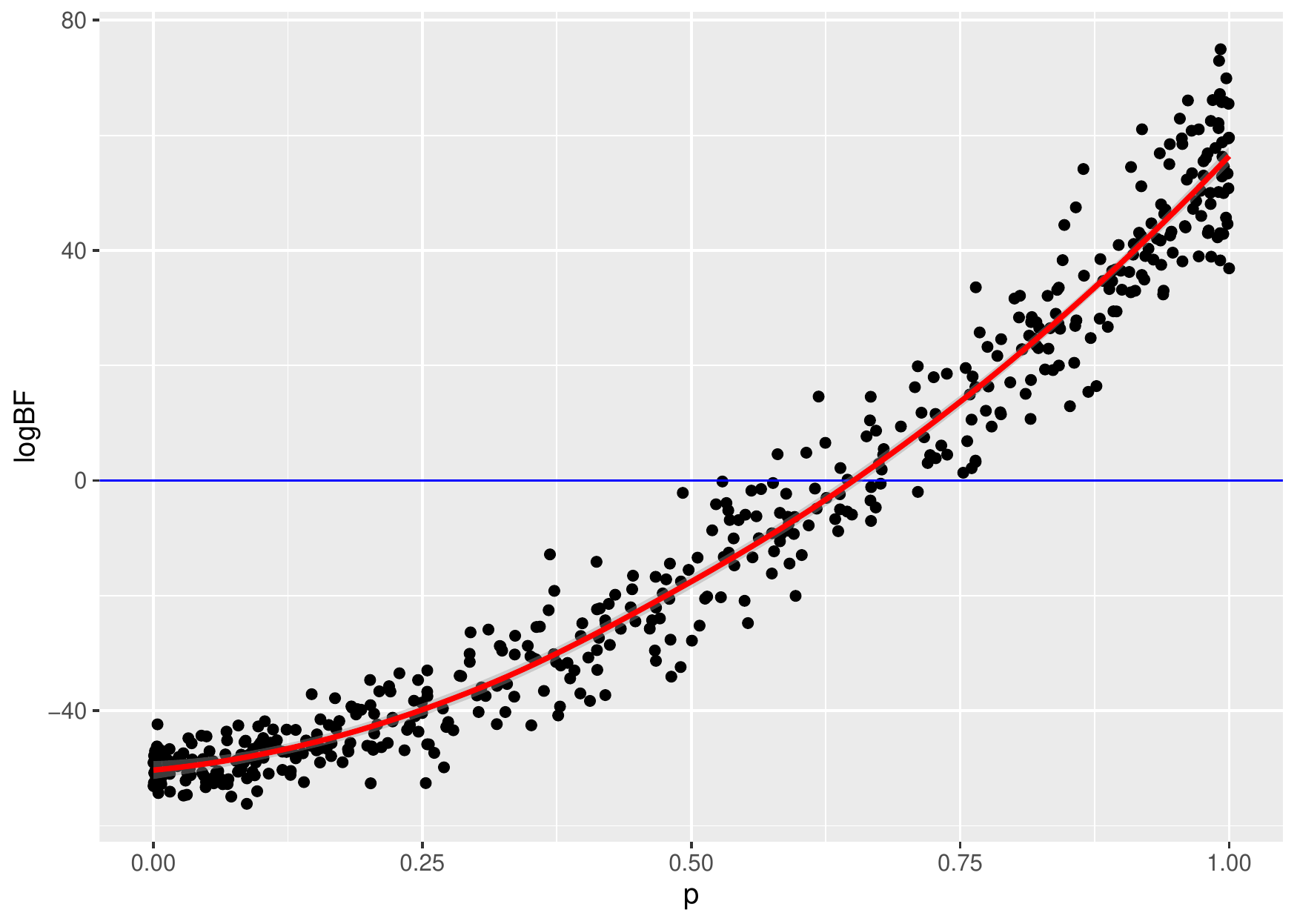} }}
    \qquad
    \subfloat[KS test log $P$-values]{{\includegraphics[width=6cm]{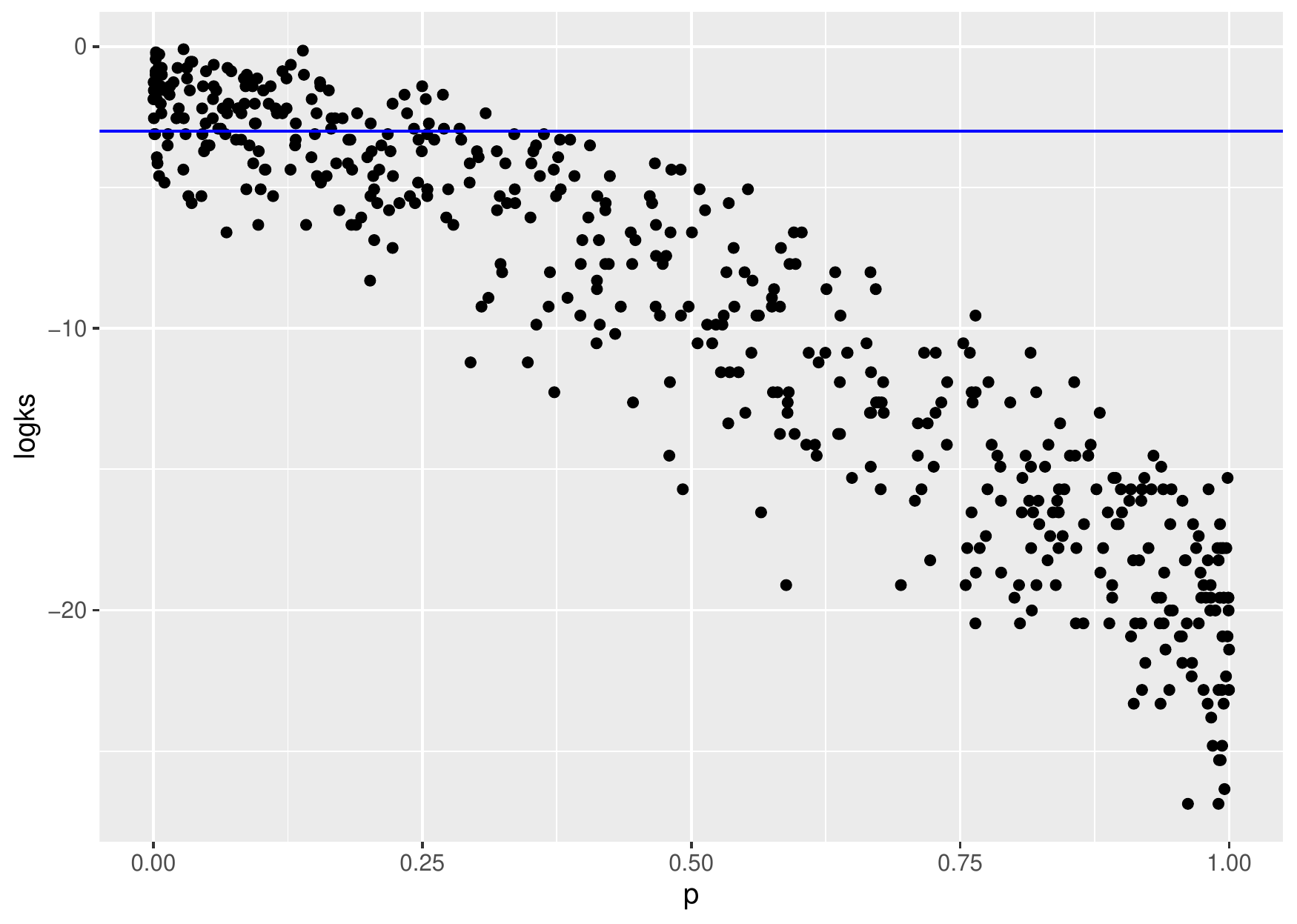} }}
    \qquad
    \subfloat[Cross-validation Bayes factors.]{{\includegraphics[width=6cm]{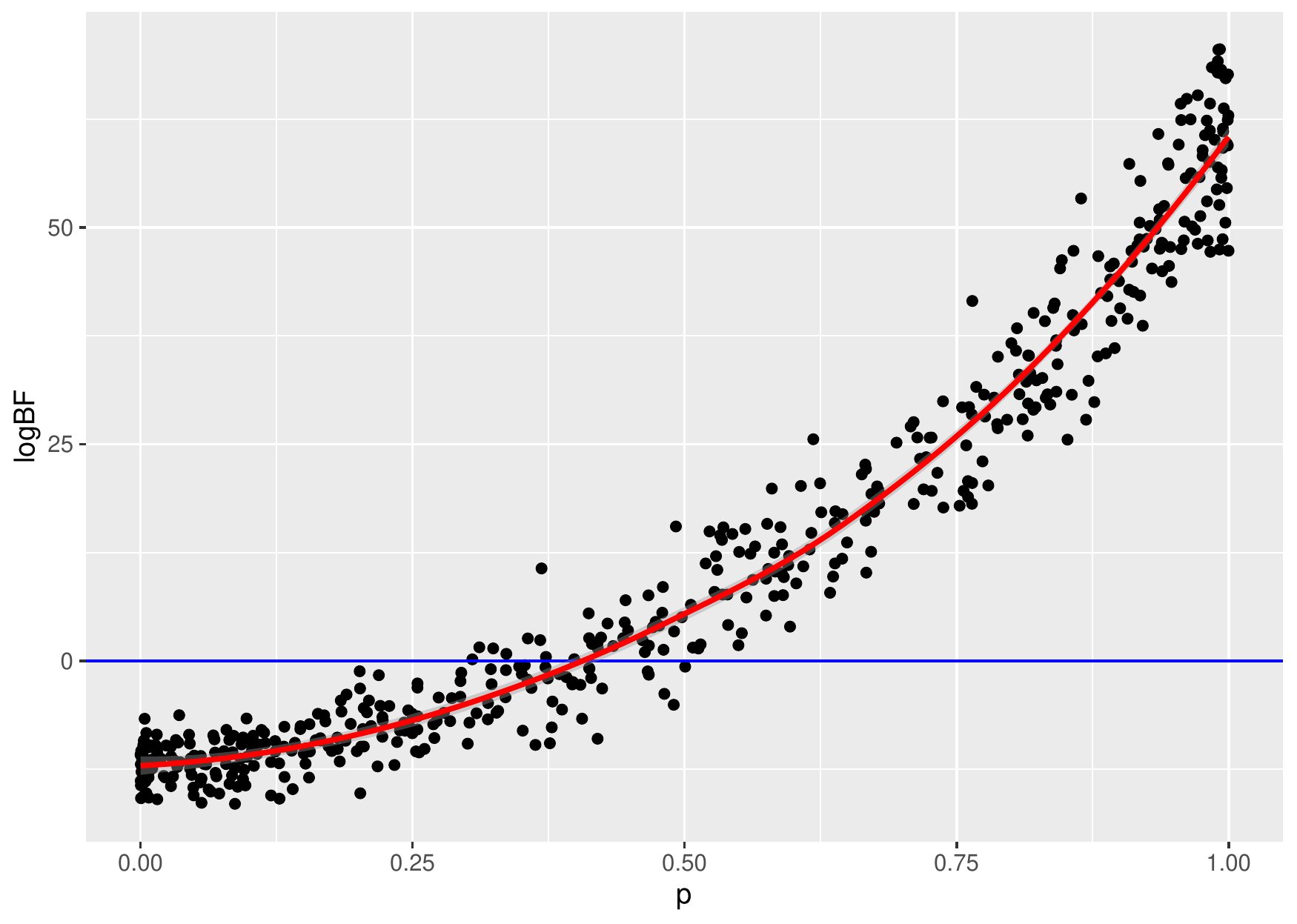} }}
    \qquad
    \subfloat[Data-reflected versions of cross-validation Bayes factors. The nonparametric estimate of standard deviation estimate is 5.06. (See text for more detail.)]
    {{\includegraphics[width=6cm]{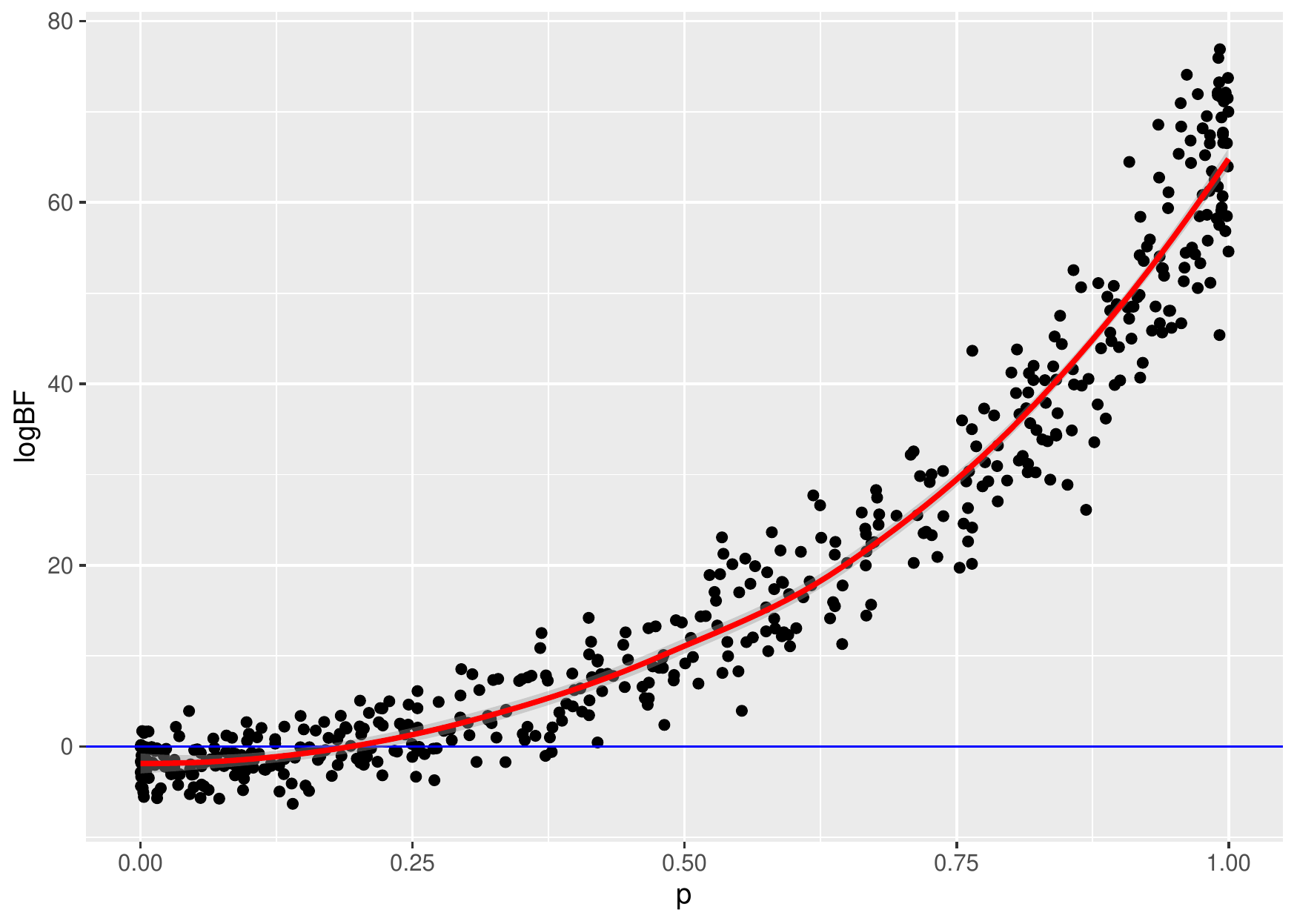}
    }}
    \caption{Comparison of tests in the case of different distributions with same finite support. Here $f$ and $g$ are $U(0,1)$ and beta$(1/2,1/2)$, respectively.}
    \label{fig:StrucDiffComp}
\end{figure}

\begin{table}
\begin{center}
\begin{tabular}{|l|c|c|c|}
\hline
&P\'olya tree: Normal&P\'olya tree: Cauchy&\\
Setting&base distribution&base distribution&CVBF\\
\hline
Scale change&3.87&4.25&3.39\\
\hline
Location shift&5.49&4.56&4.21\\
\hline
Different tail behavior&4.31&2.89&3.56\\
\hline
Finite support&5.78&6.48&4.45\\
\hline
\end{tabular}
\caption{ {\it Estimated standard deviations of log-Bayes factors.}}
\label{sds} 
\end{center}
\end{table}

Figures 5-8 provide a comparison of CVBFs, P\'olya tree Bayes factors and $P$-values of KS tests, where each of the three quantities is on the (natural) log scale. 
A blue line represents what can be considered the cut-off between evidence favoring one hypothesis over the other. In the case of the KS test the line is at $\text{log}(.05)$, which is often considered to be the largest level of significance for which the null hypothesis should be rejected. For the Bayes factors, the blue lines are at 0, as a log-Bayes factor less than 0 favors the null hypothesis (of equal densities) and one greater than 0 favors the alternative.

To perform the P\'olya Tree test, specification of a
precision parameter and a base distribution are
required. \cite{hanson2006inference} recommends 
centering and scaling the data and using a standard normal
distribution as the base distribution. However, doing so turns out not
to be efficient when the underlying distribution has sufficiently
heavy tails. In the location shift case, for example, the
Cauchy density turns out to be far better suited 
for the base distribution than the normal density since the original
density is itself Cauchy. \cite{holmes2015two} provide a data-driven procedure for choosing a base distribution, but do not show that the resulting Bayes factor is consistent under the alternative. For this reason, as well as the fact that the conditional procedure is more computationally intensive, we specify a base distribution, and study its effects. We computed P\'olya tree Bayes factors for both normal and Cauchy base distributions in each of the four settings.

The following remarks are in order concerning the simulations under alternatives. To facilitate the discussion, we refer to the P\'olya tree methodology based on normal and Cauchy base distributions as PN and PC, respectively.

\bi

\item In general, the average P\'olya tree log-Bayes factor tends to increase as the mixing parameter increases, but, depending on the base distribution, it does not rise above 0 until the mixing parameter is relatively large. This entails that a frequentist strategy would sometimes be needed to ensure good power for a test based on the P\'olya tree methodology.

\item The behavior of the P\'olya tree Bayes factors definitely depends on the base distribution used. Worse yet, the PN Bayes factors performed very poorly when at least one of $f$ and $g$ was Cauchy. In contrast, the performance of CVBFs based on the kernel $K_0$ was always comparable to or better than that of both PN and PC.

\item The PC Bayes factors performed reasonably well in all four settings, suggesting that the Cauchy might be a good default choice of base distribution.  However, the performance of PC in the finite support setting was not nearly as good as that of PN and CVBF. Also, PC Bayes factors were usually more variable than the PN Bayes factors, suggesting that PC may be less powerful in a frequentist sense than PN. 

\item Taken together, the last two remarks suggest that $K_0$ is at least a very good candidate for default kernel choice in the CVBF methodology, whereas identifying a good default base distribution in the P\'olya tree methodology is more of an open question.

\item Comparing the KS tests with the Bayes tests is not easily done since the interpretation of the $P$-value is so much different than that of a Bayes factor. However, in the case where $f$ and $g$ differed in terms of tail behavior (Figure  \ref{fig:TailDiffComp}), the behavior of PC and CVBF was clearly better than that of the KS test.

\item Table \ref{sds} provides estimated standard deviations for the log-Bayes factors (assuming homoscedasticity over the mixing parameter $p$).  Each standard deviation is the square root of the following nonparametric variance estimate: $\sum_{i = 2}^{500} (b_{i} - b_{i-1})^2/1000$, where $b_i$ is the log-Bayes factor at $p_{(i)}$, $i=1,\ldots,500$, and $p_{(1)}<p_{(2)}<\cdots<p_{(500)}$ denote the ordered values of the randomly selected mixing parameters. In most cases, the variability of the log-CVBF values is smaller than that of the P\'olya tree log-Bayes factors.  Importantly, the smaller variability of CVBF is understated as only thirty splits of each data set were used. Recall that in the null case the standard deviations of log-CVBF were about 3/4 of the standard deviations of the P\'olya tree log-Bayes factors. So, in addition to often providing more evidence in favor of the correct hypothesis, CVBF appears, in most cases, to be more stable than the P\'olya tree method.

\item In the finite support case (Figure \ref{fig:StrucDiffComp}), there are two sets of CVBF results. One set is obtained as in the other three cases, and the other set uses methodology that adjusts kernel estimates for boundary effects.  Kernel estimates are known to have large bias near a boundary when the density is positive at the boundary. To deal with the boundary bias we used a data reflection technique. First, we applied the $-\log$ transformation to each of the training data values (which yields exponential data when the underlying distribution is $U(0,1)$). We then reflected these transformed data across the $y$-axis, and constructed a  kernel density estimate from a combination of the original and reflected data. Doing so improved the behavior of the log-Bayes factors remarkably. In general this illustrates another point: methods of improving the kernel density estimate may be applied, and doing so can positively impact the log Bayes factors. While this also seems to be the case when choosing which base distribution should be used for the P\'olya tree test, we assert that there is more literature on modification of kernel density estimates than for fine-tuning P\'olya tree base distributions. See, for example, \cite{KLvE}, \cite{CH}, and \cite{BRZ}. 
  
\ei

\section{Data analysis}\label{data}

We now apply our method to the Higgs boson data set that is 
available from the UCL Machine Learning repository. The
original data set is quite large. It has 29 columns and 11 million
rows. The first column is a 0-1 variable indicating 
whether the data are noise or signal, and the rest of the columns
are variables used for distinguishing between noise and signal.  
The 2nd to 22nd columns consist of predictors, while the 23rd to 29th
columns are functions of columns 2 to 22 that are typically used for
classification. We will illustrate our methodology by applying it to the data in columns 23 and 29.

\begin{figure}[H]

\begin{center}

\includegraphics[height=3in]{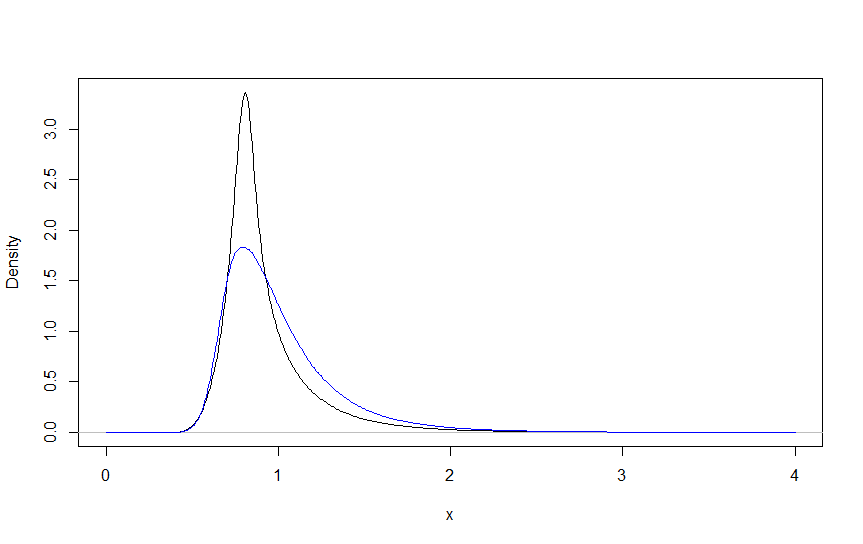}

\caption{\it Kernel density estimates for column 29 of the Higgs boson data. The blue curve is for the noise data and the black for signal.}  

\label{fig:KDE29full}

\end{center}

\end{figure}

Figure \ref{fig:KDE29full} provides KDEs for the signal and noise data in column 29. These estimates use all 11,000,000 rows of the data set. Since the two estimates are quite different one would hope that application of our methodology to even "moderate" sized samples from the two groups would tend to support the hypothesis of unequal densities. To investigate this question, we randomly selected 20,000 rows of the column 29 data.  This resulted in $m=9543$ and $n=10,457$ noise and signal 
observations, respectively. Training set sizes
$r=s=1000,2000,3000,4000,5000$ were considered, and $CVBF$ was computed for 20 different
random data splits at each $r$. The 
resulting values are provided in 
Figure \ref{fig:CVBFdifftrain}. Regardless of the training set size,
the evidence in favor of a difference between signal and noise
distributions is overwhelming. Interestingly, the results are in
agreement with expression (\ref{logLRalt}), which suggests that when
the alternative 
is true, the weight of evidence in favor of the alternative tends to
decrease with an increase in training set sizes. These results are consistent with those from the P\'olya tree and KS tests. The log-Bayes factor for the P\'olya tree test was 234.7242, while the $P$-value from the KS test was essentially 0.

\begin{figure}[H]

\begin{center}

\includegraphics[height=3in]{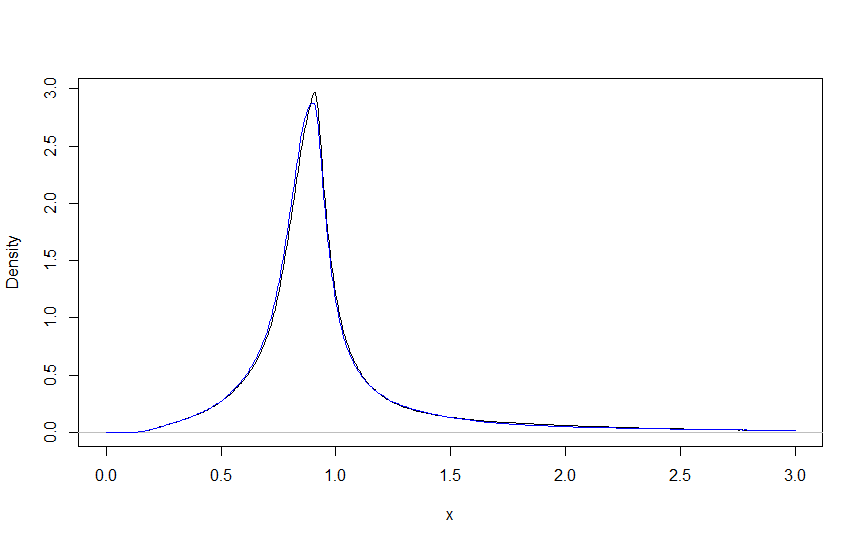}

\caption{\it Kernel density estimates for column 23 of the Higgs boson data. The blue curve is for the noise data and the black for signal.}  

\label{fig:KDE23full}

\end{center}

\end{figure}

\begin{figure}[H]

\begin{center}

{\includegraphics[height=3in]{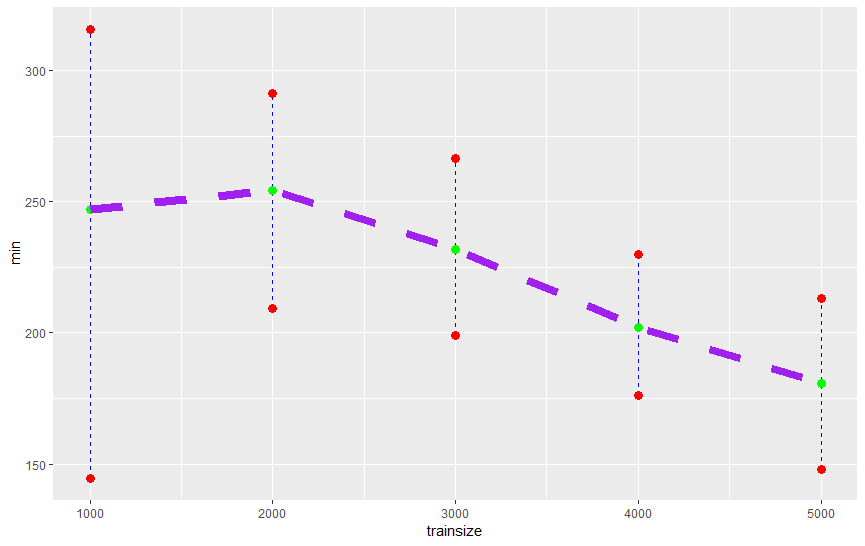}}

\caption{\it Values of log-CVBF computed from column 29 of the Higgs boson
  data. The lines connect the averages of log-CVBF at different training
  set sizes.}  

\label{fig:CVBFdifftrain}

\end{center}

\end{figure} 

\begin{figure}[H]

\begin{center}

{\includegraphics[height=3in]{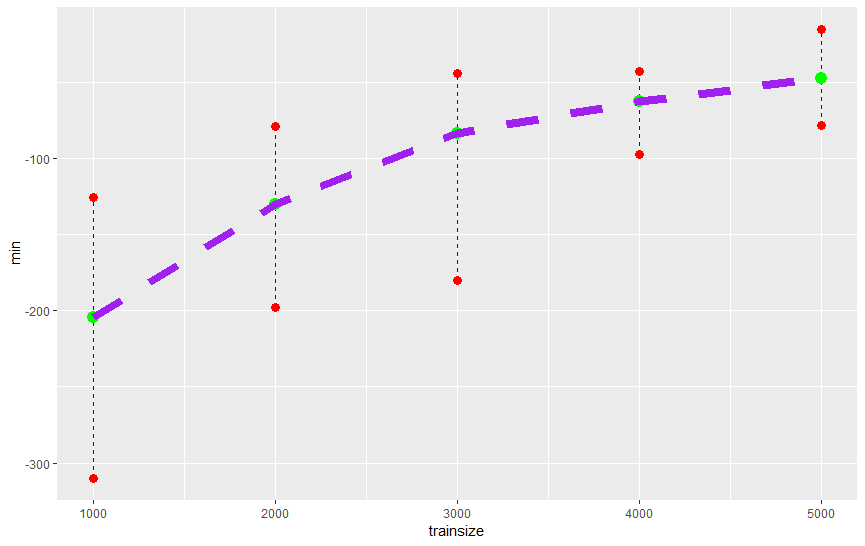}}

\caption{\it Values of log-CVBF computed from column 23 of the Higgs boson
  data. The lines connect the averages of log-CVBF at different training
  set sizes.}

\label{fig:CVBFdifftrain2}

\end{center}
\end{figure}

We now consider the column
23 data. Figure \ref{fig:KDE23full} shows signal and noise KDEs computed from all 11 million data.  Since the difference between the two estimates is extremely small, it 
would not be surprising if Bayes factors based on a small subset of the data support the hypothesis of equal densities. Proceeding exactly as in the case of column 29 data produced the results in Figure \ref{fig:CVBFdifftrain2}, where it is seen that all the
log-Bayes factors computed were smaller than $-15$. This figure shows
that the average of log-$CVBF$ increases as the training set size
increases. Based on expression (\ref{nulllimit}), this agrees with
what we expect under the null hypothesis of equal distributions.
We also considered the use of the P\'olya tree and K-S tests on
the column 23 data. These methods reach the same basic conclusion as
our procedure. The log-Bayes factor of the P\'olya tree test was -263.6514, which is
in strong favor of the null hypothesis, and the $P$-value of the KS
test is 0.1115, which is typically viewed as not small enough to
reject the null hypothesis. We also tried the approach outlined at the end of Section \ref{tv}. We used a prior that assigned equal probability to each of the following training set sizes for both noise and signal observations: 1000, 1190, 1415, 1684, 2003, 2383, 2834, 3372, 4011 and 4772, values that increase approximately linearly on a log scale.  One training set for each of the ten sizes was randomly selected, and the Bayes factor (\ref{AverOverTrains}) was calculated. This procedure was repeated ten times, leading to ten different Bayes factors of the form (\ref{AverOverTrains}),  the average of which was -53.22.

\begin{figure}[H]

\begin{center}

{\includegraphics[height=3in]{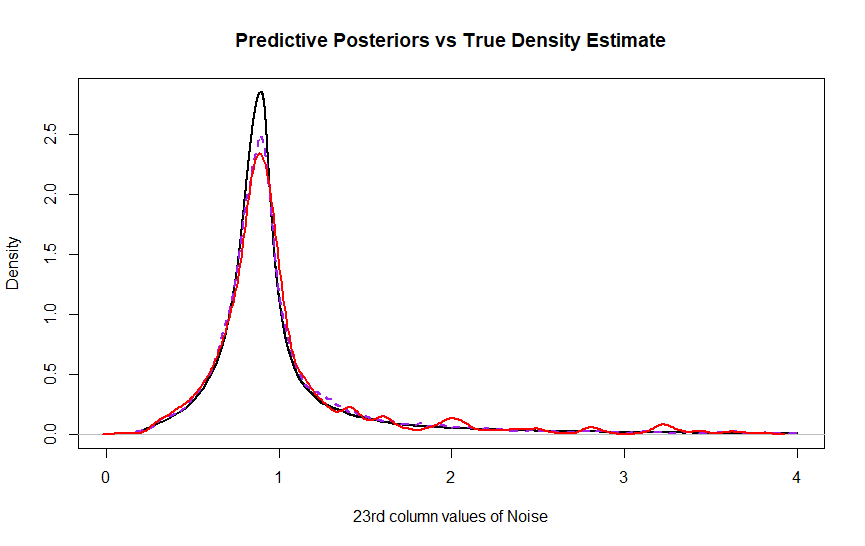}}

\caption{\it Plots of posterior predictive densities for the
  column 23 Higgs boson noise data. The black line
  is a KDE based on all 5,170,877 noise data. Because of the size of
  the data set, we regard this KDE as the truth. The red curve is the
  posterior predictive density corresponding to the P\'olya tree
  method, and in purple is a cross-validation posterior predictive
  density.} 

\label{fig:PredictivePosteriors23Noise}

\end{center}
\end{figure}

Although our focus has been on testing, it is of some interest to
  see how our cross-validatory methodology compares with P\'olya trees
  in {\it estimating} the underlying densities. To this end we compute
posterior predictive densities for the column 23 noise data using both our
methodology and P\'olya trees. Denote the first 9543 values of the
column 23 noise data by $x_1,\ld,x_{9543}$. In regard to our
cross-validation method, the posterior distribution of the bandwidth is 
$$
\pi(h|\x_V)\propto \pi(h)\prod_{i=5001}^{9543}\hat f(x_i|h,\x_T),
$$
where $\x_T=(x_1,\ld,x_{5000})$ are the training data and
$\x_V=(x_{5001},\ld,x_{9543})$ the validation data. We drew 250
values, $h_1,\ld,h_{250}$, from $\pi(\,\cdot\,|\x_V)$ using an
independence-sampler version of Metropolis-Hastings with a normal
proposal distribution that was a close match to the posterior. Our
approximation $p_{\rm pred}$ of the posterior predictive density  was 
$$
p_{\rm pred}(x)=\frac{1}{250}\sum_{i=1}^{250}\hat f(x|h_i,\x_T),
$$ 
which is plotted in Figure \ref{fig:PredictivePosteriors23Noise} along
with the P\'olya tree posterior predictive density.

The P\'olya tree method is well known for producing spurious 
  modes in the posterior predictive density \citep{hanson2006inference}.  This is
  evident in Figure \ref{fig:PredictivePosteriors23Noise} in the right
  tail of the P\'olya tree density where the data are relatively
  sparse. The cross-validation density does not seem to be subject to
  this problem, at least not to the same degree. In any event, the
  cross-validatory method has produced at least as good an
  estimate of the underlying density without the necessity
  of a complex prior distribution. This illustrates a basic
  tenet of this paper: one may devise a good Bayesian nonparametric
  procedure that does not depend on a large number of parameters and
  the attendant prior specification.
  
\section{Discussion}

We have proposed and studied a non-parametric, Bayesian two-sample test for checking equality of distributions. The methodology uses cross-validation Bayes factors (CVBFs), defining kernel density estimate models from training data and then calculating a Bayes factor from validation data.  It is advocated that a CVBF be used in genuine Bayesian fashion, i.e., by interpreting it as the relative odds of the two hypotheses.  This is in contrast to the proposal of \cite{holmes2015two}, who evaluate a P\'olya tree Bayes factor in frequentist fashion using a permutation test.  We argue that the CVBF is Bayes factor consistent under both hypotheses, and under the null hypothesis it converges in probability to 0 at an exceptionally fast rate. We provide a supplementary R package that calculates CVBFs and the P\'olya tree Bayes factor of \cite{holmes2015two}, assuming that a base distribution is supplied for the latter method.

Depending on how many data splits are utilized, calculating the average of several values of $CVBF$ can be slower than calculating the P\'olya tree Bayes factor. In particular, CVBF computations do not scale as well with the size of the data set as do those of the P\'olya tree procedure. This is mainly due to the fact that maximizing (with respect to bandwidth) the likelihood of kernel density estimates  can be time consuming. A future research problem involves the attempt to speed up the test by utilizing techniques that speed bandwidth selection. There is a great deal of information regarding how to speed up KDE calculations. Binning the data and utilizing bagging to select a bandwidth are methods that can perhaps be used to speed up CVBF calculations.

Finally, it is worth mentioning that the idea of CVBF can be generalized in a fairly straightforward fashion to deal with other inference problems, including comparison of multivariate densities, comparison of more than two densities and comparison of regression functions.

\section*{Appendix}
 Here we derive $\widehat H$, as defined at the bottom of p.~8. Let $\hat f_h$ be a KDE based on data $Z_1,\ld,Z_k$ and kernel $K$, and for arbitrary scalar quantities $u_1,\ld,u_\ell$ define $L_1$ as follows:
 $$
 L_1(h)=\prod_{j=1}^\ell\hat f_h(u_j).
 $$
 Then $L_1$ has the same structure as $L_0$ in Section \ref{lapapp}, and it suffices to consider
 \begin{equation}\label{App1}
 \frac{\partial^2}{\partial h^2}\log L_1(h)=\sum_{j=1}^{\ell}\left[\hat f_h(u_j)\frac{\partial^2}{\partial h^2}\hat f_h(u_j)-\left(\frac{\partial}{\partial h}\hat f_h(u_j)\right)^2\right]/\hat f_h^2(u_j).
 \end{equation}
 We have
 \begin{equation}\label{App2}
 \frac{\partial}{\partial h}\hat f_h(u_j)=-\frac{1}{h}\left[\hat f_h(u_j)-\hat e_h(u_j)\right],
 \end{equation}
 where $\hat e_h$ is a kernel estimator based on data $Z_1,\ld,Z_k$ and kernel 
 $J(u)=-uK'(u)$. Note that $\hat e_h$ is a ``legitimate" kernel estimator in that $\int_{-\infty}^\infty J(u)\,du=1$ and $\int_{-\infty}^\infty uJ(u)\,du=0$.
 
 Now,
 \begin{equation}\label{App3}
 \frac{\partial^2}{\partial h^2}\hat f_h(u_j)=-\frac{1}{h}\left[2\frac{\partial}{\partial h}\hat f_h(u_j)-\frac{\partial}{\partial h}\hat e_h(u_j)\right],
 \end{equation}
 and 
 $$
 \frac{\partial}{\partial h}\hat e_h(u_j)=-\frac{1}{h}\left[\hat e_h(u_j)-\hat g_h(u_j)\right],
 $$
 where $\hat g_h$ is a kernel estimator based on data $Z_1,\ld,Z_k$ and kernel $L(u)=-uJ'(u)$. As before, $\hat g_h$ is a legitimate, i.e., consistent, density estimator. Substitution of (\ref{App2}) and (\ref{App3}) into (\ref{App1}) leads to a readily computable expression for $\widehat H$.

\bibliographystyle{chicago}
\bibliography{compdens}
\end{document}